\documentclass[]{aa}  

\usepackage{graphicx,nicefrac}
\usepackage{txfonts}
\usepackage{hyperref}
\usepackage{color}
\usepackage{pdflscape}
\usepackage{multirow}
\usepackage[capitalise]{cleveref}

\DeclareRobustCommand{\ion}[2]{\textup{#1\,\textsc{\lowercase{#2}}}}
\DeclareRobustCommand{\teff}{T_{\mathrm{eff}}}
\DeclareRobustCommand{\logg}{\log g}
\DeclareRobustCommand{\mh}{\mathrm{[M/H]}}
\DeclareRobustCommand{\feh}{\mathrm{[Fe/H]}}
\DeclareRobustCommand{\vmic}{\varv_\mathrm{mic}}

\DeclareRobustCommand{\ispec}{\texttt{iSpec} }
\DeclareRobustCommand{\kms}{\mathrm{km s}^{-1}}
\DeclareRobustCommand{\vr}{v_\mathrm{r}}
\DeclareRobustCommand{\rgc}{R_{\mathrm{GC}}}

\DeclareRobustCommand{\gaiaedr3}{\emph{Gaia} EDR3}
\mathchardef\mhyphen="2D

\begin{document}

   \title{Unraveling UBC 274: a morphological, kinematical and chemical analysis of a disrupting open cluster\thanks{Thanks to observations at MIKE on the 6.5m Clay Magellan Telescope in Las Campanas Observatory}}
   \titlerunning{Unraveling the disrupted open cluster UBC 274}
   \author{L. Casamiquela\inst{1,2} \and 
    J. Olivares\inst{3,4} \and Y. Tarricq\inst{2} \and S. Ferrone\inst{1} \and C. Soubiran\inst{2} \and P. Jofré\inst{5} \and P. di Matteo\inst{1} \and F. Espinoza-Rojas\inst{6} \and A. Castro-Ginard\inst{7} \and D. de Brito Silva\inst{5} \and J. Chanamé\inst{6}}
   \institute{GEPI, Observatoire de Paris, PSL Research University, CNRS, Sorbonne Paris Cité, 5 place Jules Janssen, 92190 Meudon, France \email{laia.casamiquela@obspm.fr}
   \and
   Laboratoire d’Astrophysique de Bordeaux, Univ. Bordeaux, CNRS, B18N, allée Geoffroy Saint-Hilaire, 33615 Pessac, France
   \and
   Instituto de Astrofísica de Canarias, E-38205 La Laguna, Tenerife, Spain.
   \and
   Universidad de La Laguna, Dpto. Astrofísica, E-38206 La Laguna, Tenerife, Spain.
   \and
   N\'ucleo de Astronom\'ia, Facultad de Ingenier\'ia y Ciencias, Universidad Diego Portales, Av. Ej\'ercito 441, Santiago, Chile
   \and
   Instituto de Astrofísica, Pontificia Universidad Católica de Chile, Av. Vicuña Mackenna 4860, 782-0436 Macul, Santiago, Chile
   \and
   Leiden Observatory, Leiden University, Niels Bohrweg 2, 2333 CA Leiden, Netherlands
   }

   \date{Received ; accepted }

  \abstract
   {Open clusters in the process of disruption help to understand the formation and evolution of the Galactic disk. The wealth and homogeneity of \emph{Gaia} data have allowed the discovery of several open clusters with signs of disruption. Detailed chemical information for these clusters is essential to study the timescales and interplay between the star formation process and cluster disruption.}
   {We do a morphological, kinematic and chemical analysis of the disrupting cluster UBC~274 (2.5 Gyr, $d=1778$ pc), with the objective of studying its global properties.}
   {We use HDBSCAN to obtain a new membership list up to 50 pc from its centre and up to magnitude $G=19$ using \gaiaedr3 data. We use high resolution and high signal-to-noise spectra to obtain atmospheric parameters of 6 giants and subgiants, and individual abundances of 18 chemical species.}
   {The cluster has a highly eccentric (0.93) component, tilted $\sim$10 deg with respect to the plane of the Galaxy, which is morphologically compatible with the result of a test-particle simulation of a disrupting cluster. Our abundance analysis  shows that the cluster has a subsolar metallicity of $\feh=-0.08\pm0.02$. Its chemical pattern is compatible with that of Ruprecht~147, of similar age but located closer to the Sun, with the remarkable exception of neutron-capture elements, which present an overabundance of $[n\mathrm{/Fe]}\sim0.1$.}
   {The cluster's elongated morphology is associated with the internal part of its tidal tail, following the expected dynamical process of disruption. We find a significant sign of mass segregation where the most massive stars appear 1.5 times more concentrated than other stars. The cluster's overabundance of neutron-capture elements can be related to the metallicity dependence of the neutron-capture yields due to the secondary nature of these elements, predicted by some models. UBC~274 presents a high chemical homogeneity at the level of $0.03$ dex in the sampled region of its tidal tails.}


   \maketitle

\section{Introduction}
The study of the open cluster (OC) population has been recently fostered, particularly after the advent of \emph{Gaia} DR2 \citep{GaiaCollaboration+2018} and the following EDR3 \citep{GaiaCollaboration+2021}. New OCs have been discovered thanks to the wealth of astrometric information from \emph{Gaia} and its exquisite precision \citep[e.g.][]{Castro-Ginard+2018,Castro-Ginard+2019,Castro-Ginard+2020, Liu+2019}. After the determination of homogeneous membership lists, several studies have revisited the general properties of the OC population, including the determination of distances, ages, reddening, morphology, 3D kinematics and orbits \citep[e.g.][]{Cantat-Gaudin+2020,Tarricq+2021}.
In parallel, some studies performed a detailed search for members in a few nearby OCs up to a large radius from the cluster centre, finding that 16 clusters have coronae and tidal structures, sometimes extending more than 100 pc from the cluster centre \citep{Meingast+2019,Roser+2019,Roser+2019b,Meingast+2021}. Recently, \citet{Tarricq+2021b} used EDR3 data to perform a larger-scale study of the 2D morphology of the OC population (up to 1.5 kpc from the Sun) and found remarkable elongations and tidal tails in the plane of the sky for 70 OCs.

The presence of coronae and elongated structures in young clusters are believed to be a result of the complexity of the star formation process \citep{Meingast+2021}, while around evolved OCs they are believed to be a sign of disruption. Gravitationally-bound systems resulting from star-formation events are believed to undergo internal and external disruption processes, which lead to mass loss during their lives \citep{Lada+2003}. Simulations usually show that these systems would lose their stars releasing them into the form of tidal tails \citep{Ernst+2011}.
As a consequence, most of these initially bound systems end up being dissolved into the field, particularly the less massive ones. This is in agreement with the fact that we observe a significant decrease in the number of known OCs with increasing age. Many processes possibly play a role in the disruption of clusters, like the birthplace of the cluster (inner disk clusters probably suffer higher disruption rates), the orbital parameters, or the interaction with giant molecular clouds \citep[][]{Wielen1985,Gieles+2016}. 

Therefore, the detailed study of clusters with signs of disruption is key to understanding the interplay among the different processes. Thanks to the nature of \emph{Gaia} data, we are now able to identify reliable members far away from the cluster cores and analyse their extended morphology and the mass segregation degree, which can provide information on the nature of the disruption. Additionally, kinematical and detailed chemical analysis can help to understand the origin of particular clusters and the conditions at their birth.

UBC~274 is an OC recently discovered by \citet{Castro-Ginard+2020}, who detected 365 members in a radius of $\sim2$ deg around the cluster centre up to magnitude $G=17$. They identified a remarkable elongation of the distribution of stars in the direction of the proper motion vector, suggesting that this cluster is in process of disruption. The cluster was further included in the sample studied by \citet{Cantat-Gaudin+2020}, who performed a search for members in a similar radius around the cluster, and reported 566 members up to magnitude $G=18$, from which 342 have a membership probability of $\geq0.7$. They also estimate its physical parameters: $\log Age=9.43$ (2.7 Gyr), $A_V=0.24$, and heliocentric distance $d=1.78$ kpc.

Furthermore, \citet{Piatti2020} used the members identified by \citet{Castro-Ginard+2020}, and found that the cluster was affected by differential reddening with values ranging $E_{(B-V)}=0.1-0.25$. They corrected the \emph{Gaia} photometry according to a reddening map and computed the cluster age from isochrone fitting: $\log Age=9.45\pm0.05$. Like this, \citet{Piatti2020} could identify the cluster binary sequence and found that the binary population has similar radial density profiles along the tidal tails as the single stars. Finally, they also investigate the internal kinematics of the cluster members and found that the cluster rotates as a solid body in the Galactic plane.

\vspace{0.5cm}

The purpose of this paper is to analyse in detail the physical properties of the cluster UBC~274, including its morphology, kinematics and its detailed chemistry thanks to follow-up high-resolution spectra. The paper is organised as follows: in Sect.~\ref{sec:membership} we extend the cluster membership determination up to magnitude 19 around 50 pc from the cluster centre, using the method developed in \citet{Tarricq+2021b}. We test the recovery and contamination rate using a synthetic cluster. We run a simulation of a disrupting cluster with the current observed 6D coordinates to unveil the disruption of this cluster (Sect.~\ref{sec:morph}), and we study the mass segregation degree that we observe with our new deeper membership (Sect.~\ref{sec:segregation}). We also use high resolution and high signal-to-noise (S/N) spectroscopic follow-up observations of several member stars to obtain detailed chemical abundances of the cluster (Sect.~\ref{sec:spec}). In Sect.~\ref{sec:discussion} we provide a discussion of the global properties of the cluster, and in Sect.~\ref{sec:conclusions} we expose the conclusions of the work.

\section{Membership, morphology and kinematics}\label{sec:memb}

Our aim is to obtain an accurate list of members at a large distance from the centre of the cluster, and test its completeness and contamination rates, to be able to analyse in-depth the cluster morphology and its link to the kinematical properties.

\subsection{Membership analysis}\label{sec:membership}

We obtained a list of members of the cluster using the same method as \cite{Tarricq+2021b} and taking the mean cluster position and velocity computed by \cite{Cantat-Gaudin+2020} as a starting point.

We did a cone search in the \gaiaedr3 archive centred on the cluster with a 50 pc radius, and up to magnitude $G=19$. We filter out the stars which have proper motions out of the 10$\sigma$ interval centred on the cluster's mean proper motion, to avoid extremely high discrepancies.
We also discarded the stars with a Renormalized Unit Weight Error (RUWE) higher than 1.4 \citep{Fabricius+2021}, because of their high probability of having a problematic astrometric solution which could lead to contamination of the member list. By rejecting sources with large RUWE values we expect to discard a considerable fraction of unresolved binaries. We have checked that stars with large RUWE values do not have a clear dependence with magnitude. The impact of the RUWE cutoff will be discussed in Sect~\ref{sec:segregation}.
We used the Python implementation of Hierarchical Density-Based Spatial Clustering of Applications with Noise (HDBSCAN) \citep{HDBSCAN, hdbscan_python} to identify the cluster members in the $(\mu_{\alpha*},\mu_{\delta},\varpi)$ space. We used the same configuration of the HDBSCAN hyperparameters as \cite{Tarricq+2021b}. To compute membership probabilities we perform 100 runs of the algorithm, using a random sampling on the proper motions and parallax according to their full covariance matrix. In this way, a star that is classified as a member in 100\% of the runs has a probability of membership of 1.

As noticed in \cite{Tarricq+2021b} the distribution of the member's probability depends on multiple factors including the heliocentric distance of the cluster, and the relative density of the cluster with respect to the surrounding field in the proper motion/parallax space. To define the members of each cluster, they chose to use 0.5 as a probability cutoff, which was considered the best compromise between contamination and recovery rates for the 389 clusters they analysed. However, UBC~274 is quite different from that sample of clusters, particularly because it is located further away than the analysed clusters in \cite{Tarricq+2021b}, and this option is not well adapted in this case. We based our probability cutoff on the analysis of the performance of the algorithm on simulations of synthetic clusters.

We generated synthetic observables of the cluster using the public python code \texttt{Amasijo}\footnote{\url{https://github.com/olivares-j/Amasijo}}. Astrometry is drawn from a gaussian distribution centred on the mean observed location and motion of the real cluster. Photometry in the \gaiaedr3 passbands is generated from the mean cluster age and metallicity, using the package \texttt{isochrones}\footnote{\url{https://github.com/timothydmorton/isochrones}} \citep{Morton2015} according to the initial mass function from \cite{2005ASSL..327...41C}. Finally, astrometric and photometric uncertainties are added to the synthetic cluster stars using \texttt{PyGaia}\footnote{\url{https://github.com/agabrown/PyGaia}}. We use the Gaia Object Generator \citep[GOG,][]{Luri+2014}, which is based on the Gaia Universe Model Snapshot \citep[GUMS,][]{Robin+2012}, to get a characterisation of the field population: we performed the same query in the \emph{Gaia} source simulation table as done for the real data in the \gaiaedr3 archive, to obtain the simulated individual sources and their uncertainties. We realised that the parallax uncertainties from the source simulation are in general overestimated with respect to the real uncertainties of \gaiaedr3 values. This fact would make us underestimate the performance of the clustering algorithm. We solved this by recomputing the uncertainties of the field simulation using \texttt{PyGaia}, in the same way, that we did for the cluster simulation.

We run the clustering algorithm to the simulated cluster and field stars in the same way as we did for the real cluster stars.
It is well known that \emph{Gaia} astrometric uncertainties heavily depend on the photometric magnitude \citep[][see fig. 7]{Lindegren+2021}. Because of this effect, any membership classifier that takes into account the astrometric uncertainties will display a strong dependence on its performance with $G$ magnitude. The bright and high precision sources will statistically display better classification properties (low contamination and high recovery rates) than the faint low-precision and therefore more entangled sources. For this reason, establishing a single probability cutoff would bias the results, and affect particularly the mass segregation analysis in Sect.~\ref{sec:segregation}. Additionally, for the particular case of this cluster, the extended area of analysis and the differential reddening makes the performance of the classifier depend on the spatial location, especially on Galactic latitude because the cluster is located quite far from the Galactic plane ($b\sim-12$). For these reasons, we computed the recovery statistics (contamination rate and true positive rate) in bins of $G$ magnitude and Galactic latitude $b$. We plot in Fig.~\ref{fig:classification} the dependence of the performance of the classifier as a function of the probability threshold used, in bins of $G$ and $b$. At each of these bins, we selected the optimum classification threshold as the probability value that maximizes the classifier's efficiency. Table~\ref{tab:thresholds} contains these optimal probability thresholds, and the expected contamination and true positive rates.

After applying the optimal probability cuts we obtain a final member list with 665 stars. This list of members represents the best compromise that we can obtain (see discussion in Appendix.~\ref{sec:appendix}). 
This membership contains 354 more stars with respect to \citet{Castro-Ginard+2020}, most of them located between $17<G<19$, since they analysed only bright sources. There are 54 stars that we do not recover from \citet{Castro-Ginard+2020} in the common magnitude range because of two reasons: i) some of them have large RUWE values and thus are discarded by our initial query, and also ii) due to the better uncertainties in EDR3 with respect to DR2 the proper motion and parallax distributions of the cluster are now narrower, and thus some of the stars are now identified as non-members. 
We obtain a similar comparison with respect to \citet[][p$>0.7$]{Cantat-Gaudin+2020}: our list contains 346 more stars, and we do not recover 23 of their members.

We attempted to recover members located up to 100 pc from the centre of the cluster. However, the performance of the algorithm was very poor in this case, for which we obtained much larger contamination visible in the colour-magnitude diagram (CMD). This is possibly because of the less clumped distribution of outer stars in the proper motions and parallax space given the large extent of the sky area, coupled with the fact that the cluster is kinematically embedded in the field distribution.

\subsection{Morphology and kinematics}\label{sec:morph}

\begin{figure}
\centering
\includegraphics[width = 0.5\textwidth]{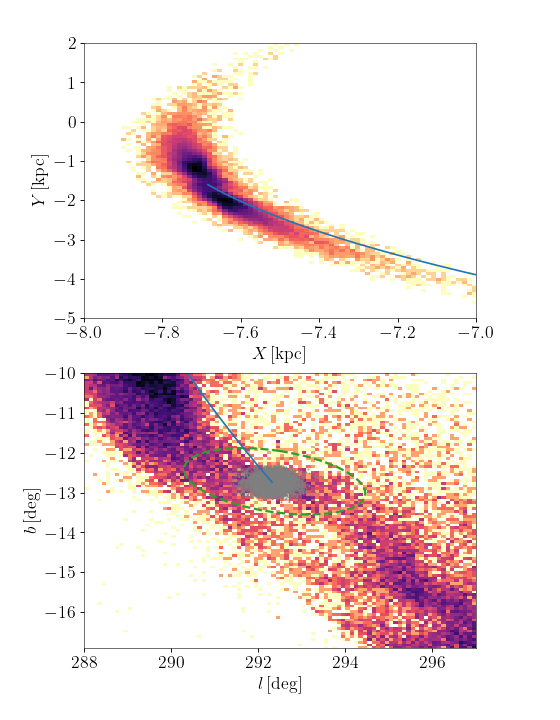}
\includegraphics[width = 0.5\textwidth]{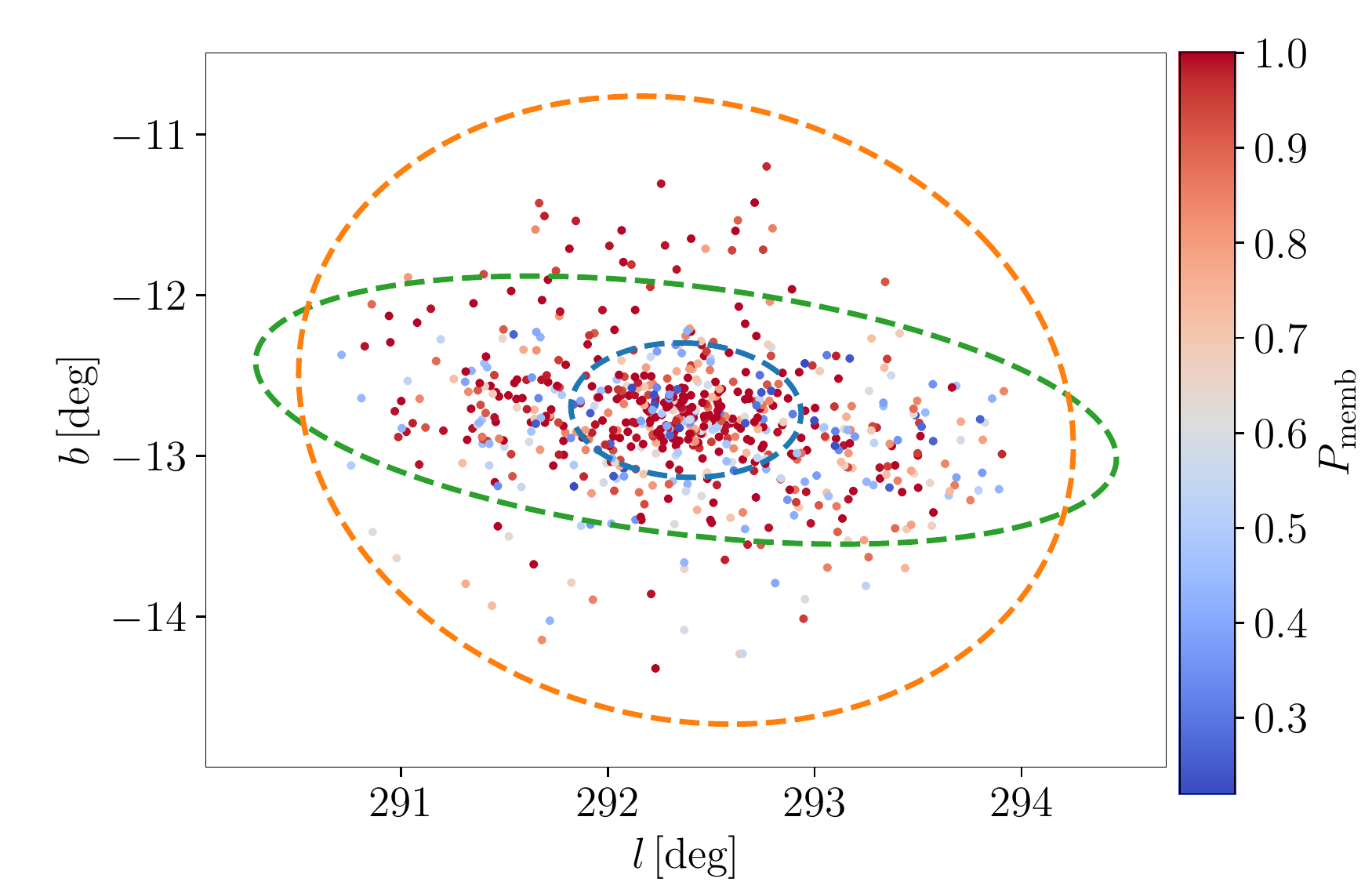}
\caption{Top and middle: ($X,Y$) and ($l,b$) distribution of the simulated escaping stars of the cluster. The colours correspond to the density of points, and the blue line indicates the orbit integrated backwards. In grey we plot the stars which are still bounded to the cluster (energy $<0$).
Bottom: ($l,b$) distribution of the real cluster members coloured by the probability of membership. The three ellipses correspond to the three Gaussian components obtained by our GMM fit in Sect.~\ref{sec:morph}, corresponding to core (blue), tidal tail (green) and halo (orange). The tidal tail ellipse is also plotted in the middle panel for comparison.}\label{fig:gmm}
\end{figure}

To better analyse the morphology of the cluster we fitted a Gaussian Mixture Model (GMM) on the projected sky distribution in Galactic coordinates. This model assumes that the whole data can be described by a combination of a finite number of Gaussians. We followed the same process as described in \citet{Tarricq+2021b} to fit the Gaussian models, obtaining the three Gaussians plotted in Fig.~\ref{fig:gmm} (bottom), and detailed in Table~\ref{tab:gmm}. Remarkably, the second component that we obtained displays a very large eccentricity of $0.93$, and is associated with the visually identified tidal tail in \citep{Castro-Ginard+2020}. This component is tilted 10 deg with respect to the horizontal plane of the Galaxy.

\begin{table}
\caption{Results of the GMM fit to three components: semi-major and semi-minor axes of the ellipses corresponding to the $3\sigma$ values of the Gaussians (as plotted in Fig.~\ref{fig:gmm}), the tilt of the ellipses with respect to the horizontal line, and the eccentricity.} \label{tab:gmm}
\centering
\setlength\tabcolsep{2.2pt}
\begin{tabular}{lcccccc}
 \hline
Component  &  Semi-major & Semi-minor & Tilt & Eccentricity\\
 & axis (deg) & axis (deg) & (deg) & \\
\hline
Core & 0.56 & 0.41 & 5  & 0.68 \\
Tail & 2.11 & 0.77 & 10 & 0.93 \\
Halo & 2.03 & 1.79 & 56 & 0.47 \\
\hline
\end{tabular}
\end{table}

Tidal tails or stellar streams are formed of escaping stars as a consequence of the interaction between the cluster initially bounded stars and the potential of their host galaxy \cite[see][]{fall1977survival}. According to theory and simulations, the cluster should release stars to the tidal structure and form an "S" shaped stream almost aligned with its orbit \citep{montuori07}. Tidal tails and streams have been historically studied in globular clusters because they are more evident due to the high number of stars that populate them, and because it is easier to identify a stream against a stellar halo background than in the plane \citep[see e.g.][]{odenkirchen01}. Even though open clusters are less massive, since they are extended objects in the Galactic gravitational field, they also experience tidal stripping. The precision, homogeneity and all-sky nature of \emph{Gaia} data has actually allowed for the systematic discovery of OC tidal tails \citep[e.g.][]{Tarricq+2021b}.

We perform a test particle simulation with the objective of comparing the expected output of a disrupting cluster of similar kinematic and physical characteristics as UBC~274. The method can be summarized as follows. We first derive the Galactocentric positions and velocities of the cluster by transforming the mean position in the sky, proper motions and radial velocity. In deriving the Galactocentric positions and velocities of the cluster, we assume the position of the Sun to be $(X_{\odot},Y_{\odot},Z_{\odot})=(-8.34, 0.0, 0.027)$ kpc \citet{reid14}, a velocity for the local standard of rest, $V_{LSR}=240$ km/s \citep{reid14}, and a peculiar velocity of the Sun with respect to the LSR,  $(U_{\odot},V_{\odot},W_{\odot})=(11.1, 12.24, 7.25)$ km/s \citep{schonrich10}.
Then we integrate the orbit of the cluster, represented as a point mass,  backward in time for 500 million years. This time corresponds approximately to two complete orbits around the Galactic centre. For the mass distribution of the Milky Way, we use one of the Galactic potential models described in \citet{pouliasis2017milky} (Model II in their paper), which represents the Galaxy as made of a stellar thin and thick discs and surrounded by a dark matter halo whose analytical form is the same adopted by \citet{Allen+1991}. We refer the reader to \citet{pouliasis2017milky} for the details of the Galactic potential. Once the position and velocity of the cluster, as to 500 Myr ago is recovered, we replace the point mass with an N-body system made of 10,000 particles, whose distribution function follows a Plummer potential, having a half-mass radius equal to 5 pc and total mass equal to 1900 $M_{\odot}$, and integrate the orbit of each particle subject to both the Galactic potential and the analytic cluster potential until today. The adopted values for the half-mass radius of the cluster and its total mass were set taking into account the observed current distribution of members, and represent a rough guess of the cluster parameters in the past. The uncertainty in these parameters will not affect the general recovered morphology of the tidal debris, but rather its density with respect to the bounded stars.

Our model does not take into account direct gravitational interactions between stars in the cluster, that is, our simulation is not an N-body simulation, but a test-particle simulation where each particle in the system is mass-less. This approach does not allow for studying the internal dynamics of the cluster, or the mass loss of the cluster over time, but is adapted to study the orientation of the tidal debris and their elongation in the disk of the Galaxy. This is a valid approach for the scope of our analysis given that internal dynamics have limited effects on tidal debris \citep{montuori07,piatti2020tidal}. We study mass segregation with other methods, as discussed in the following section.

The ($X,Y$) and ($l,b$) distribution of the simulated stream are plotted in the top panels of Fig.~\ref{fig:gmm}. Comparing the distribution of the cluster stars with that of the simulation, we see that the direction of elongation observed in the data is compatible with the internal part of the S shape predicted by the simulation. The slight tilt that we observe in the elongated component of the GMM is well aligned with the internal part of the simulated stream. In contrast, we do not observe the high density stellar structure that forms the external stream, which is aligned with the orbit. This is because the cone search where we performed the membership does not cover this wide area. We tried enlarging the radius of the cone search up to 100 pc from the cluster centre, but as explained in Sect.~\ref{sec:memb}, our method showed poor performance, with very large contamination rates.

The resemblance between the real morphology and the simulation shows that UBC~274 is indeed in the process of disruption following the expected dynamical process described by our simulation. Other studies found the elongation of several nearby clusters to be in the direction of the plane of the Galaxy \citep{Meingast+2021,Hu+2021}, which is compatible with what we see in UBC~274, and what is expected from circular disk orbits, as shown by our simulation.

We remark the fact that our simulation predicts several overdensities in the streams, which are unfortunately outside of the covered observational area. Overdensities in tidal streams of globular clusters have been attributed to epicyclic motions of the stars escaping from a disrupting cluster \citep[e.g.][]{Kupper+2010}. More precise data from the future \emph{Gaia} data releases and coupled with ground-based radial velocities will probably allow sampling of this region of the cluster and study if this phenomenon is present in disrupting open clusters.

\subsection{Mass segregation}\label{sec:segregation}
Another consequence of dynamical evolution in clusters is mass segregation. This is explained by the equipartition of kinetic energy via two-body relaxation: massive stars within a cluster move towards its centre whereas lighter stars move towards its outskirts \citep{delafuentemarcos1996}. This process highly depends on the initial mass function of the cluster and its total mass, but also on other more subtle effects from stellar evolution, which changes directly the ratio of heavy to light stars.

\begin{figure}
\centering
\includegraphics[width = 0.5\textwidth]{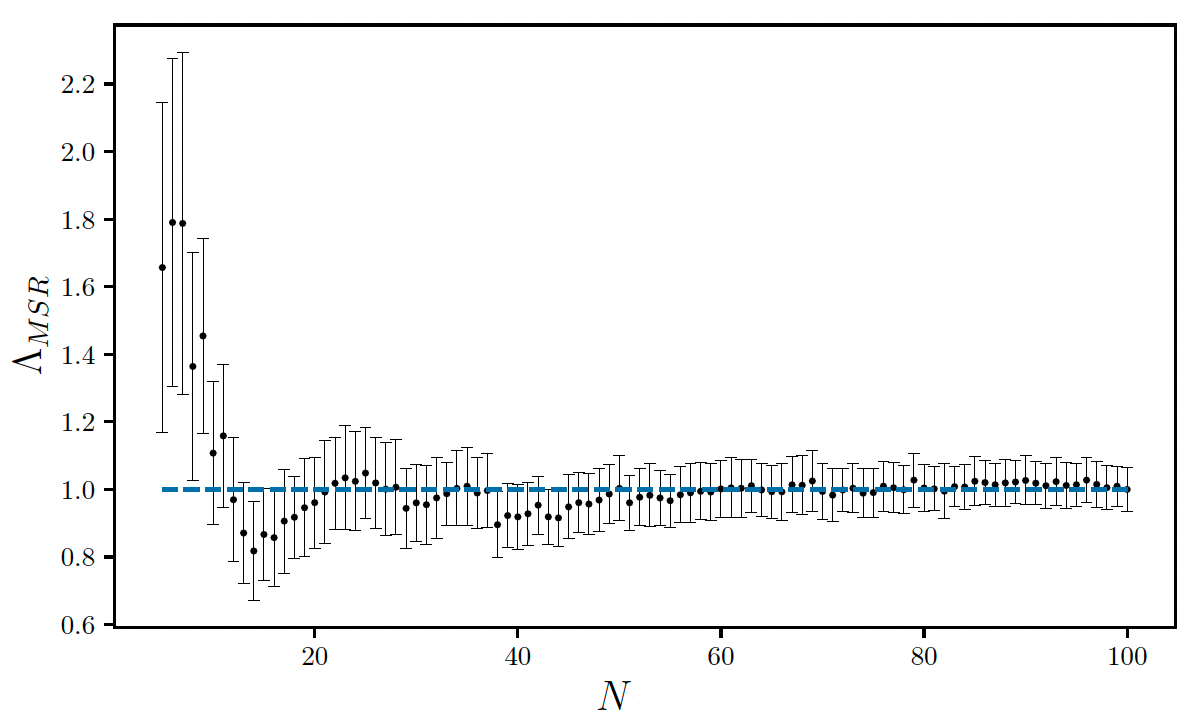}
\caption{Mass segregation ratio $\Lambda_{MSR}$ as a function of the number of stars $N$. The horizontal line marks the $\Lambda_{MSR}=1$ (no mass segregation).}\label{fig:masssegregation}
\end{figure}

We have used the method by \citet[][]{Allison+2009} to measure the degree of mass segregation. It consists in comparing the length of the minimum spanning tree (MST) of massive stars with that of random stars. If there is mass segregation, the MST length of the most massive stars will be shorter than that of random stars. Then we define the mass segregation ratio of the N most massive stars as:

\begin{equation}
    \Lambda_{MSR}\,(N) = \frac{\langle l_{N,\mathrm{random}}\rangle}{l_{N,\mathrm{massive}}} \pm \frac{\sigma_{N,\mathrm{random}}}{l_{N,\mathrm{massive}}}
\end{equation}

where $l_{N,\mathrm{random}}$ is the average length of the MST of N randomly chosen stars and $l_{N,\mathrm{massive}}$ is the length of the MST of the N most massive stars. The average length $\langle l_{N,\mathrm{random}}\rangle$ was calculated over 100 iterations where at each iteration we draw a different subsample of random stars allowing us to calculate at the same time $\sigma_{N,\mathrm{random}}$, the standard deviation of the length of the MST of these N stars. This is the same strategy used in \citet{Tarricq+2021b}. We use the observed $G$ magnitude as a proxy for the mass. We computed this ratio for increasing values of $N$ obtaining the values plotted in Fig.~\ref{fig:masssegregation}. 

If for a particular $N$, the $\Lambda_{MSR}$ is larger than 1, it means that the $N$ most massive stars are more concentrated than $N$ randomly chosen stars. In this case, we see that the 10 most massive stars have a significant ratio of mass segregation $\Lambda_{MSR}\,(N=10)\sim1.5$, which means that they are on average 1.5 times closer than randomly chosen stars. For larger $N$ this ratio rapidly decreases to 1 and remains stable, indicating no mass segregation in the majority of the range of masses. We have checked that this result is independent of the probability threshold used to construct the member list. Adopting a single and more restrictive membership probability cutoff of 0.5 and 0.7 to avoid any possible contamination, we obtain compatible results with those in Fig.~\ref{fig:masssegregation}, where the points only vary among uncertainties.

Compared with the distribution of mass segregation $\Lambda_{MSR}\,(N=10)$ computed for more than 300 clusters in \citet{Tarricq+2021b}, UBC~274 shows a similar mass segregation degree to the clusters of its age. \citet{Olivares+2019} also found strong evidence of mass segregation for the cluster Ruprecht~147, with similar age as UBC~274.

A significant mass segregation for this cluster contrasts with the findings of \citet{Piatti2020}, who studied the spatial distribution of binaries vs single stars in UBC~274, finding that there is no overall difference between the two. They argue that if internal dynamics in a cluster was driven by the two-body relaxation, one would also expect binary stars to be more centrally concentrated with respect to single stars. This made them conclude that the disruption swept out any sign of mass segregation. Here we see a significant degree of mass segregation but obtained using intrinsically massive stars (basically dominated by AGB and giants), instead of binaries.
Our method presents a more generalised evaluation of mass segregation than using binaries since one would expect binaries to be more centrally located than single stars only for a given magnitude bin. In other words, a low mass binary is not expected to be more centrally located than a single more massive star. Moreover, the identification of binaries is very sensitive to the uncertainties in differential extinction values across the cluster region. Thus, overall, the signal of mass segregation could be hidden when evaluating it using only binaries.

We do not attempt to identify and analyse the distribution of binaries in this cluster since our membership procedure tends to discard a significant fraction of unresolved binaries (by imposing a maximum RUWE value in Sect.~\ref{sec:membership}), thus we do not expect to have a complete census of the binary systems in this cluster. However, this does not have an impact on our conclusions because the mentioned filter removes unresolved binaries regardless of their total mass, thus, statistically, it should have no impact when comparing the MST of bright vs faint stars. The error bars in our Fig.~\ref{fig:masssegregation} partly account for this effect.

The different conclusions obtained by this work and \citet{Piatti2020} may come from the different range of masses that the two procedures take into account. They compared binaries and non-binaries in the main sequence of the cluster, where binaries can be best identified, while we find the mass segregation signature only for AGBs and giants. For less massive stars (subgiants, turnoff and main sequence), in Fig.~\ref{fig:masssegregation} we see an abrupt decrease of the $\Lambda_{MSR}\,(N)$ reaching rapidly values of 1 for $N>15$. Thus, it is possible that the highest mass binaries that they identified are less massive than the stars that populate the AGB, and therefore the identified binaries would not reflect the mass segregation signature.

\section{Spectroscopic analysis}\label{sec:spec}
We did spectroscopic follow-up observations of eight stars, with the purpose of deriving high precision chemical abundances for reliable members.

\subsection{Data}\label{sec:data}

We carried out observations of the spectroscopic targets with the Magellan Inamori Kyocera Echelle (MIKE) spectrograph \citep{Bernstein+2003} on the 6.5m Clay Magellan Telescope at Las Campanas Observatory in a 2 night run in January 2021. The nominal spectral resolution of the instrument is $R = 83,000$ and $R = 65,000$ in the blue and red arms of MIKE, respectively, and we used the slit of 0.7''. The orders were extracted with the CarnegiePython MIKE pipeline\footnote{\url{ http://code.obs.carnegiescience.edu/mike}} and the  continuum normalization and merge of the orders were performed with PyRAF.

\begin{figure*}
\centering
\includegraphics[width = \textwidth]{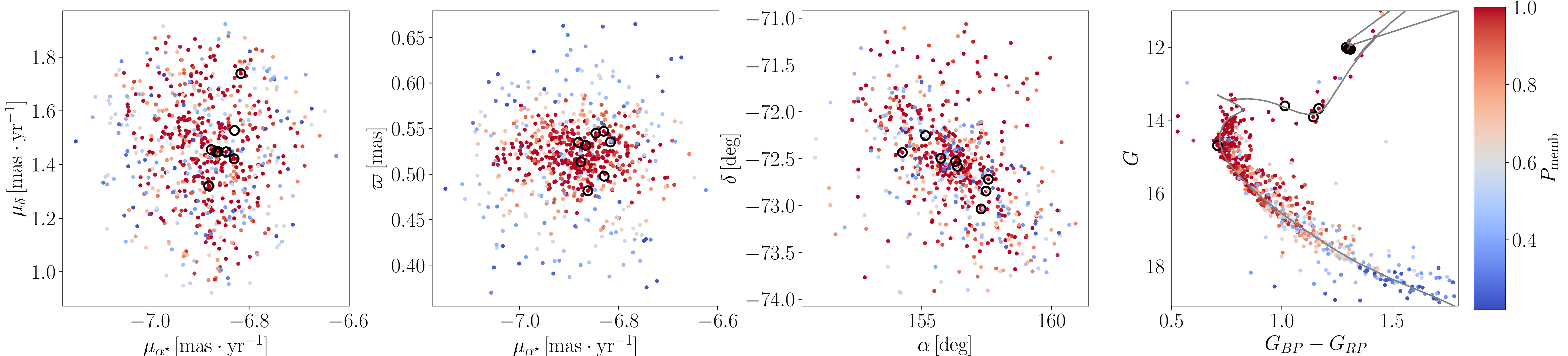}
\caption{Members of the cluster obtained in Sec.~\ref{sec:memb} coloured by membership probability. Spectroscopic targets marked with black circles. From left to right: proper motions distribution, proper motion-parallax distribution, spatial distribution in equatorial coordinates, and CMD in \gaiaedr3 passbands. The isochrone from the PARSEC database \citep{Bressan+2012} of the cluster age and metallicity (2.5 Gyr and $\mh=-0.1$as derived in Sect.~\ref{sec:cluster} and shifted by $d = 1778$ pc and $E_{(B-V)}=0.12$) is overplotted in the CMD to guide the eye.}\label{fig:targets}
\end{figure*}

By the time the observations were made EDR3 was not available, so we did our target selection using the high probability members of the cluster obtained by \citet{Cantat-Gaudin+2020}, based on \emph{Gaia} DR2. We chose stars brighter than magnitude $G=14$, given the technical constraints and the aimed S/N ($\gtrsim100$). We excluded the stars with inconsistent radial velocity from \emph{Gaia} DR2 compared to the cluster mean value obtained \citep{Castro-Ginard+2020} from 13 members $-22.92 \pm 1.26\,\kms$. Also, we tried to cover as much as possible the cluster RA-DEC extension.
In total, we obtained 18 spectra of 8 stars with exposure times between 300 s and 1000 s. Most of the targets were observed several times, and those spectra were later coadded to reach the maximum S/N. A summary of the spectroscopic targets and the observation details is found in Table~\ref{tab:observations}. The EDR3 proper motions, parallax, spatial distribution and CMD of the cluster members and targets observed are plotted in Fig.~\ref{fig:targets}.

We checked that all spectroscopic targets have the same source id in \gaiaedr3. There is one star (source id: 5230056103835667200) which gets rejected in our membership procedure (Sect.~\ref{sec:memb}), because it has a RUWE of 1.55, out of the imposed limits. Its proper motions and parallax are well within the distribution of the rest of the cluster, and it corresponds to the hottest subgiant in the CMD plot. We discuss this star later in Sect.~\ref{sec:rvs}.

\begin{table}
\caption{Details of the spectroscopic observations. Stars are identified using the Gaia DR2/EDR3 source id, and we include their magnitude in $G$ band, the probability of membership (PM$_{CG}$) obtained by \citet{Cantat-Gaudin+2020}, and that derived in Sect.~\ref{sec:memb} (PM$_{here}$). For each star, we indicate the number of spectra acquired (Num) and the signal to noise of the coadded spectrum. Flags are indicated in the text of Sect.~\ref{sec:rvs}.}\label{tab:observations}
\centering
\setlength\tabcolsep{2.2pt}
\begin{tabular}{lcccccc}
 \hline
source\_id  &  $G_{mag}$ & PM$_{CG}$ & PM$_{here}$ & Num & S/N & Flag\\
\hline
5229199446838976512 & 12.1 & 0.9 & 1.0 & 1 & 122 &  \\ 
5229225938195844736 & 12.1 & 0.9 & 1.0 & 2 & 152 &  \\ 
5229228339075497088 & 12.0 & 1.0 & 1.0 & 2 & 156 &  \\ 
5230035724210643584 & 12.1 & 0.9 & 1.0 & 1 & 129 &  \\ 
5229986765883047680 & 13.7 & 1.0 & 1.0 & 4 & 127 &  \\ 
5230056103835667200 & 13.6 & 0.9 &  -  & 4 & 268 & 1 \\ 
5229998791791507968 & 13.9 & 1.0 & 1.0 & 2 & 113 & 2 \\ 
5229979275459995392 & 14.7 & 0.9 & 1.0 & 2 & 185 & 3 \\ 
\hline
\end{tabular}
\end{table}

To perform the analysis of the spectra we used the public spectroscopic software \ispec \citep{BlancoCuaresma+2014,BlancoCuaresma2019}. We made use of a similar pipeline to the one used in \citet{Casamiquela+2020}. 
As a first step, the spectra were cut to the wavelength range of the selected spectral lines (480-680 nm). After the determination of radial velocities using a cross-correlation algorithm, telluric regions and emission lines due to cosmic rays were masked, and a further normalization was performed using a median and maximum filtering. Then, atmospheric parameters and abundances were computed as explained in Sect.~\ref{sec:AP},\ref{sec:chemistry}.

\subsection{Radial velocities}\label{sec:rvs}
Among the stars considered as members in the study by \citet{Cantat-Gaudin+2020}, 19 stars have radial velocity measurements from \emph{Gaia} DR2 RVS. Two of them were considered as non-members/spectroscopic binaries in \citep{Castro-Ginard+2020} with incompatible the radial velocities with respect to the rest of the cluster and were discarded from our starting list.

\begin{figure}
\centering
\includegraphics[width=0.5\textwidth]{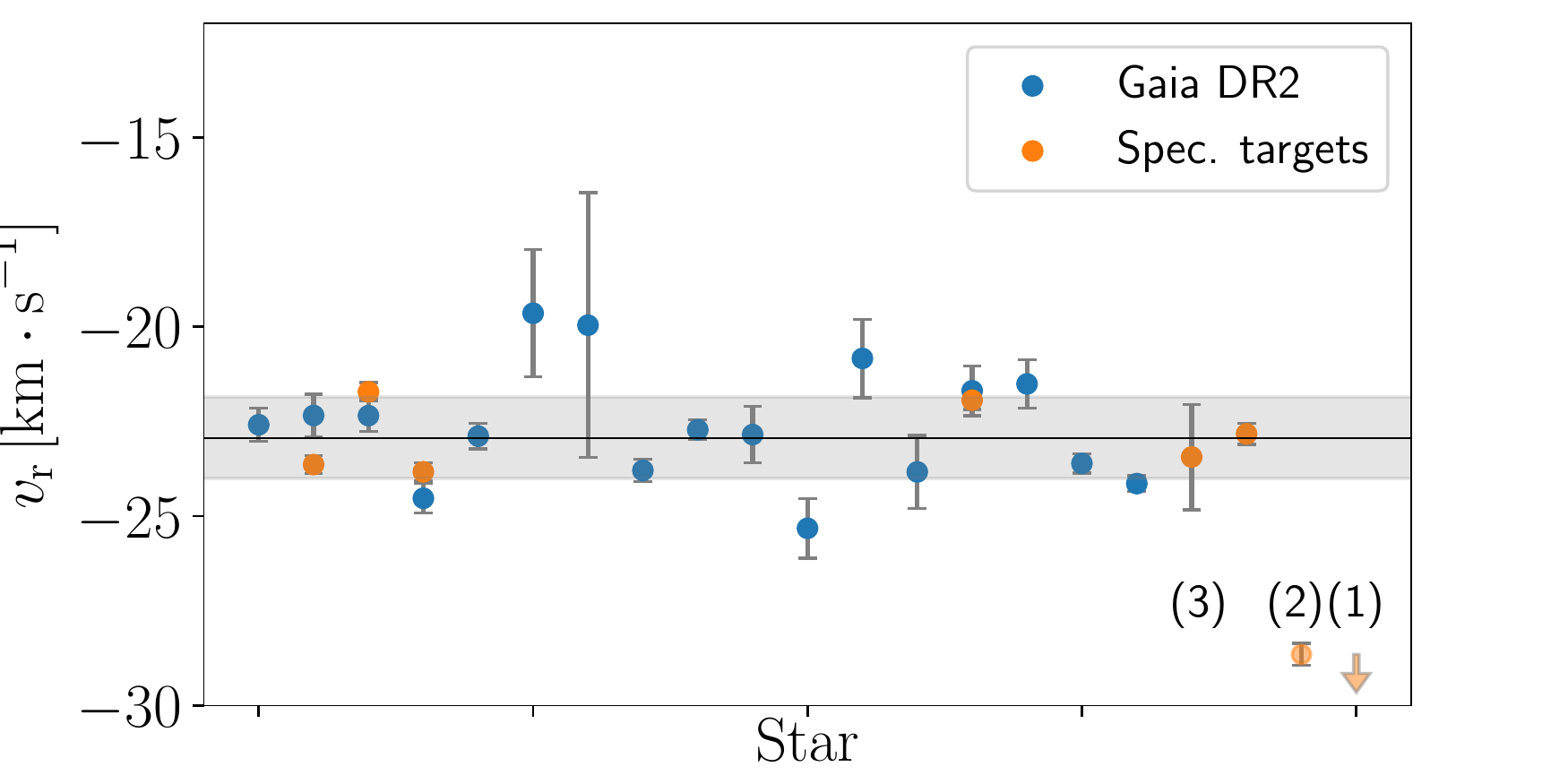}
\caption{Radial velocity values measured by \emph{Gaia} DR2 RVS considered as members in \citep{Castro-Ginard+2020} (blue), and values measured in the spectroscopic analysis (orange). The arrow (flag 1) indicates the discrepant star mentioned in the text with a radial velocity $-52.71\pm1.95\,\kms$. The weighted mean and standard deviation is plotted with a solid line and a shadowed region.}\label{fig:rvs}
\end{figure}

The $\vr$ values of the 17 compatible members measured from \emph{Gaia} DR2 RVS are plotted in Fig.~\ref{fig:rvs}. We also plot the radial velocity values obtained by our pipeline for the eight spectroscopic targets.

The four spectroscopic targets in common with the RVS members show that both sets of radial velocities are in good agreement below the $\sim1\,\kms$ level, without any systematic offset. Of the four spectroscopic targets not in common with RVS, two are compatible with the rest of the stars. The two rightmost stars plotted in Fig.~\ref{fig:rvs} have a $\vr$ not compatible at more than $3\sigma$ with respect to the mean of the rest of the stars. We highlight three particular cases:

\begin{itemize}
    \item Gaia DR2 5230056103835667200 (flagged as 1 in the plot), has a $\vr=-52.71\pm1.95\,\kms$ and large width of the spectral features which causes a large FWHM of the cross-correlation function (40$\,\kms$, see also Fig.~\ref{fig:AP}). This is probably an indication of rotation and/or binarity. Additionally, as explained in Sect.~\ref{sec:data} this star is not classified as a member in our membership study because of its large RUWE, which in turn can be a sign of binarity \citep[][]{Penoyre+2021}. Altogether, this points towards this star being a binary or a non-member.
    \item Gaia DR2 5229998791791507968 (flagged as 2 in the plot), has a $\vr=-28.65\pm0.29\,\kms$, and shows a similar FWHM compared to the other stars. Its relatively small difference in $\vr$ with respect to the cluster, and the fact that its atmospheric parameters and chemical abundances are in agreement with the cluster (see Sect.~\ref{sec:spec}) makes us think that it is a spectroscopic binary without contamination of the companion star to the spectrum.
    \item Gaia DR2 5229979275459995392 (flagged as 3 in the plot) has a compatible radial velocity, but larger uncertainties, owing to its large FWHM. For its position near the turnoff, it could be due to high rotation. We do not discard it for the radial velocity computation.
\end{itemize}

In Fig.~\ref{fig:rvs} we also plot the weighted mean and dispersion of the cluster radial velocity ($\vr=-22.9\pm1.1\,\kms$), rejecting the two rightmost stars. For the four stars in common with \emph{Gaia} DR2 RVS we first perform a weighted mean of the two $\vr$ values.

\subsection{Atmospheric parameters}\label{sec:AP}
We used the spectral synthesis fitting method to compute atmospheric parameters ($\teff$, $\logg$, $\mh$, $[\alpha\mathrm{/M}]$, and $\vmic$). The fitting was done by comparing the uncertainty-weighted fluxes of a set of observed features with a synthetic spectrum. Atmospheric parameters were varied until the convergence of a least-squares algorithm was reached. The whole process made use of the local thermodynamic equilibrium (LTE) radiative transfer code SPECTRUM \citep{Gray+1994} and the MARCS atmospheric models \citep{Gustafsson2008}. We used version 6 of the \emph{Gaia}-ESO line list \citep{Heiter+2021}.

\begin{figure}
\centering
\includegraphics[width=0.5\textwidth]{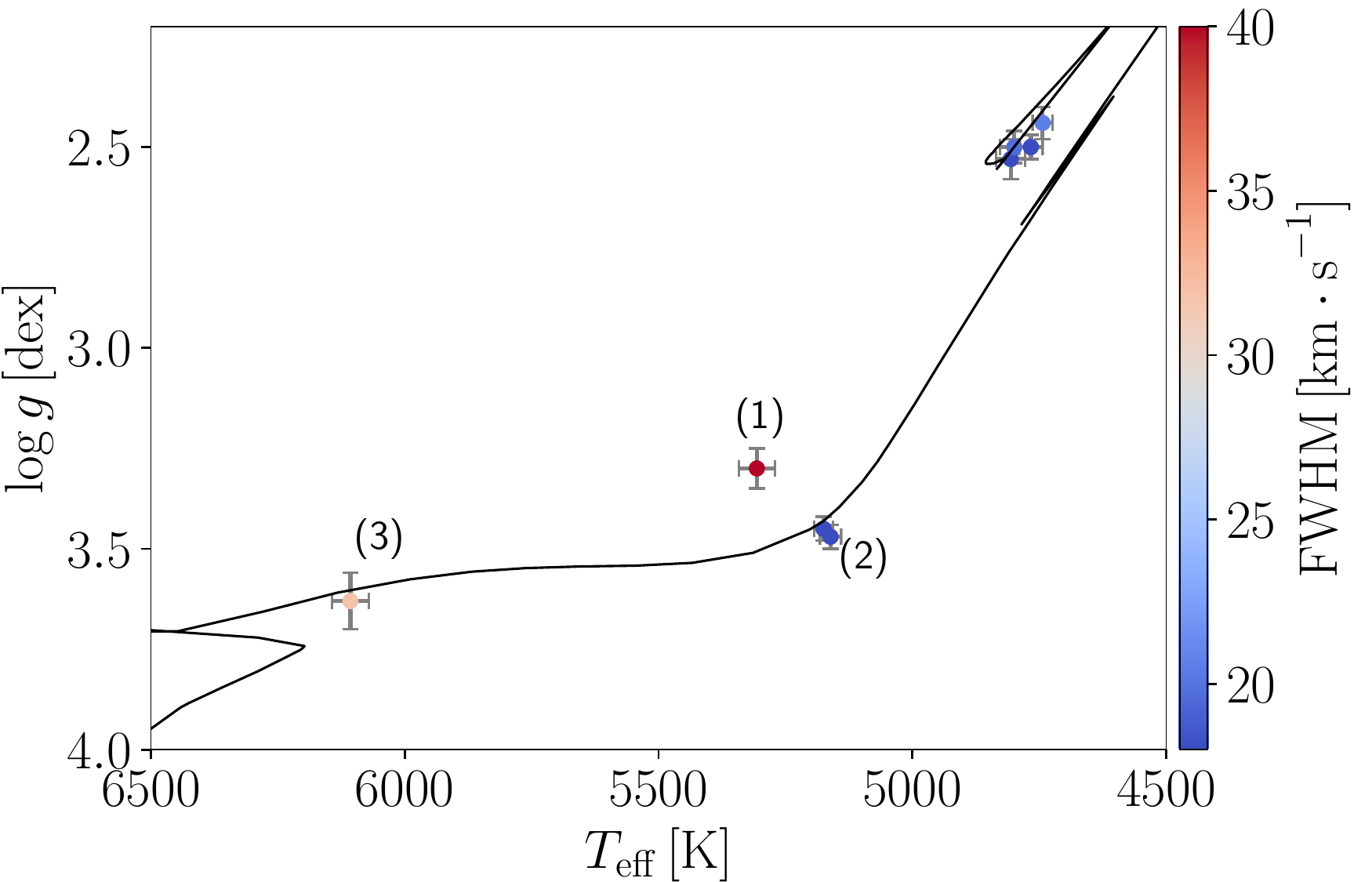}
\caption{Kiel diagram of the spectroscopic targets coloured by the FWHM of the spectral lines. Isochrone of the cluster age and metallicity is overplotted. Flagged stars as described in the text are identified. The flagged stars discussed in the text are indicated.}\label{fig:AP}
\end{figure}

We plot in Fig.~\ref{fig:AP} the results of the $\teff$ and $\logg$ derived by our pipeline. In general, the results agree very well with the overplotted theoretical isochrone of age $2.5$ Gyr and $\mh=-0.1$ (Sect.~\ref{sec:chemprop}). 

We highlight in the plot the flagged stars in Sect.~\ref{sec:rvs}. The star (1) does not fit well with the isochrone, presents very high FWHM, discrepant radial velocity, and is not retained by our membership because of large RUWE (with high chances of being a spectroscopic binary or a non-member).
The star (2) corresponds to the faintest subgiant in the CMD. Though it is discrepant in radial velocity, the mean FWHM of the lines is very similar to that of the other subgiant, and the spectroscopic determinations of the atmospheric parameters fit well the isochrone. Finally, the star (3), corresponds to the upper main-sequence star in the CMD. We find it is a turnoff star judging by its atmospheric parameters and the age of the cluster. As mentioned in Sect.~\ref{sec:rvs}, its radial velocity agrees well with the cluster, though with larger uncertainties. Its FWHM$\sim30\,\kms$ is a sign of rotation.

In Table~\ref{tab:APs} we include the radial velocities and atmospheric parameters computed for the spectroscopic targets. We can see that the metallicity of the stars flagged as (1) and (3) is much smaller than the other targets. Given the spectroscopic peculiarities of these two stars, we consider that their chemical characterisation is not as accurate as for the other targets, which have more consistent metallicity values. For this reason, we decide to exclude stars (1) and (3) from the chemical analysis in the following subsection. In the light of the results, we consider that star (2) can be a spectroscopic binary without contamination of the companion star, for this reason, we keep it for the chemical abundance analysis.

\begin{table*}
\caption{Radial velocities and atmospheric parameters of the stars analysed.}\label{tab:APs}
\centering
\setlength\tabcolsep{2pt}
\begin{tabular}{lccccccccc}
 \hline
source\_id & $\vr\,[\kms]$ & $\teff\,[\mathrm{K}]$ & $\logg\,[\mathrm{dex}]$ & $\vmic\,[\kms]$ & $\mh$ & Flag & Comment \\
 \hline
5229199446838976512  & $-23.64 \pm 0.24$ & $4806\pm28$ & $2.53\pm0.05$ & $1.50\pm0.03$ & $-0.11\pm0.02$ &   &  \\
5229225938195844736  & $-21.72 \pm 0.24$ & $4766\pm23$ & $2.50\pm0.03$ & $1.46\pm0.02$ & $-0.10\pm0.02$ &   &  \\
5229228339075497088  & $-23.83 \pm 0.24$ & $4743\pm19$ & $2.44\pm0.04$ & $1.45\pm0.02$ & $-0.15\pm0.02$ &   &  \\
5230035724210643584  & $-21.94 \pm 0.25$ & $4799\pm28$ & $2.50\pm0.04$ & $1.48\pm0.03$ & $-0.14\pm0.02$ &   &  \\
5229986765883047680  & $-22.83 \pm 0.28$ & $5175\pm18$ & $3.45\pm0.03$ & $1.28\pm0.03$ & $-0.05\pm0.01$ &   &  \\
5230056103835667200  & $-52.71 \pm 1.95$ & $5306\pm35$ & $3.30\pm0.05$ & $0.81\pm0.06$ & $-0.30\pm0.03$ & 1 & rotation/NM/SB \\
5229998791791507968  & $-28.65 \pm 0.29$ & $5161\pm20$ & $3.47\pm0.03$ & $1.27\pm0.03$ & $-0.09\pm0.01$ & 2 & SB \\
5229979275459995392  & $-23.44 \pm 1.39$ & $6107\pm36$ & $3.63\pm0.07$ & $1.70\pm0.07$ & $-0.26\pm0.02$ & 3 & rotation \\
\hline
\end{tabular}

\small \flushleft SB: spectroscopic binary, NM: non-member
\end{table*}

\subsection{Chemical abundances}\label{sec:chemistry}
Individual absolute chemical abundances per spectrum were obtained using the atmospheric parameters fixed to the resulting values of the previous step. We used the same radiative transfer code (SPECTRUM), model atmospheres (MARCS), line list, and fitting algorithm. Hyperfine structure splitting and isotopic shifts were taken into account for the elements: V I, Mn I, Co I, Cu I, Ba II, La II, PrII, and Nd II, following \citet{Heiter+2015b,Heiter+2021}.
Neither non-LTE nor 3D corrections were taken into account in the computation of the abundances.

We additionally analysed a Solar spectrum (from the asteroid Vesta) of S/N$=530$ acquired with MIKE in the same configuration as the cluster stars. This was used to compute bracket abundances in a non-differential manner. We refer to bracket abundances as $[\mathrm{X/H}]=A_X-A_{X,\odot}$, with $A_X$ being the absolute abundance of the star for the element X, and $A_{X,\odot}$ the absolute abundance of the Sun. Absolute abundances are defined as $A_X = \log\left(\frac{N_X}{N_H}\right) + 12$, where $N_X$ and $N_H$ are the number of absorbers of the element X and of hydrogen, respectively.

In parallel, we computed strictly line-by-line differential abundances with the same strategy as \citet{Casamiquela+2021}, see more details of this procedure in their sect~3. In brief, differential abundances were calculated line by line, subtracting the absolute abundance values of a chosen reference star. As reference star we used the high S/N spectrum of the red clump star from the Hyades HD~27371 analysed in \citet[][$\teff=4975 \pm 12$ K, $\logg = 2.83 \pm 0.03$]{Casamiquela+2020}. For its atmospheric parameters, this star is considered an analogue to the giants of UBC~274. Later, to compute Solar-scaled abundances, we used the differential analysis of the Hyades Solar analogue HD~28344 which uses the Sun as reference star \citep[][$\teff=5957\pm32$ K, $\logg=4.49\pm0.03$, $\feh=0.12\pm0.05$]{Casamiquela+2020}. 

\begin{figure*}
\centering
\includegraphics[width=\textwidth]{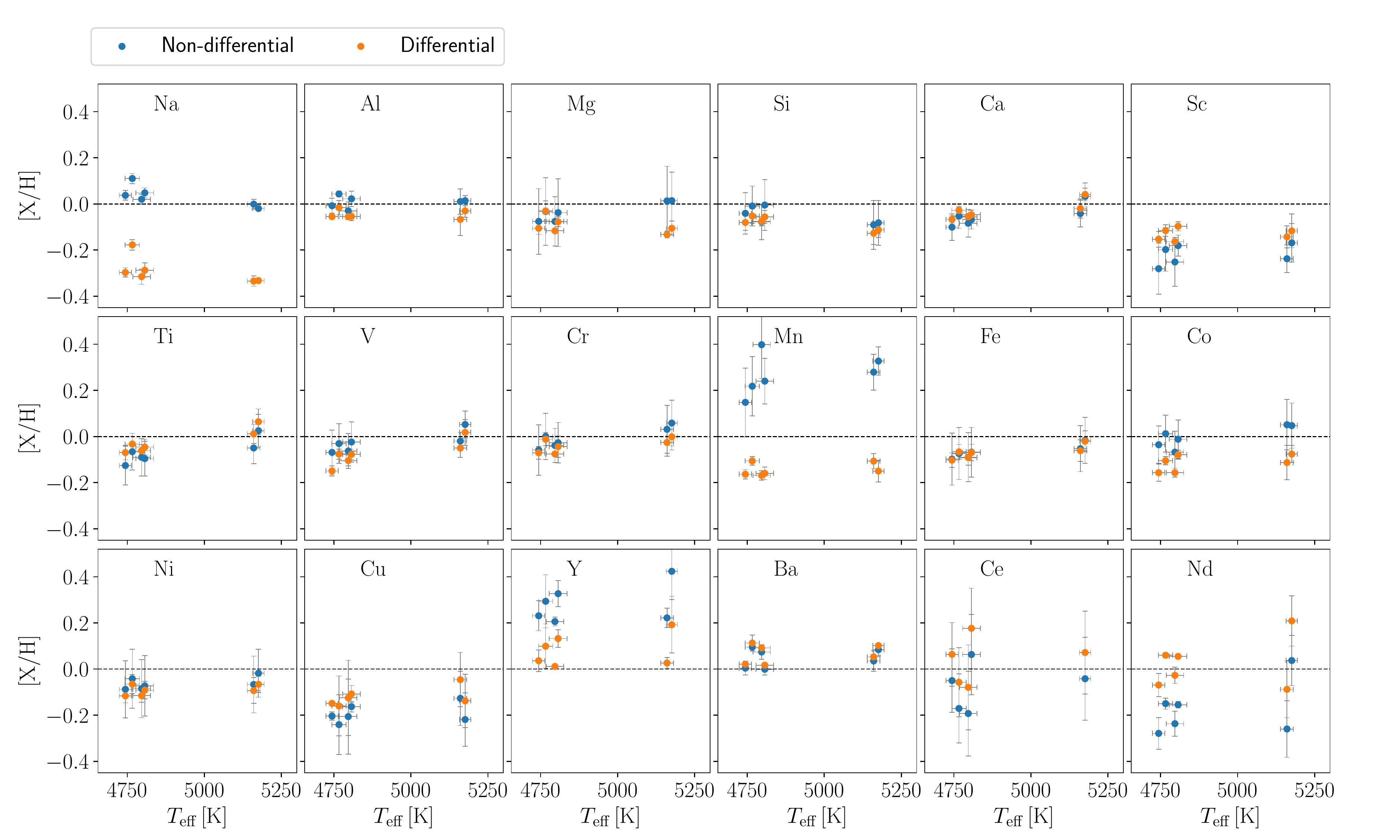}
\caption{Chemical Solar-scaled abundances of the four giants and two subgiants of the cluster computed in a differential (orange) and non-differential manner (blue).}\label{fig:chemab}
\end{figure*}

We plot in Fig.~\ref{fig:chemab} the computed Solar-scaled abundances for the four red clump stars and the two subgiants of the cluster. The non-differential bracket abundances are plotted in blue in Fig.~\ref{fig:chemab}, and the strictly differential values transformed to the Solar scale are plotted in orange.

\begin{figure}
\centering
\includegraphics[width=0.5\textwidth]{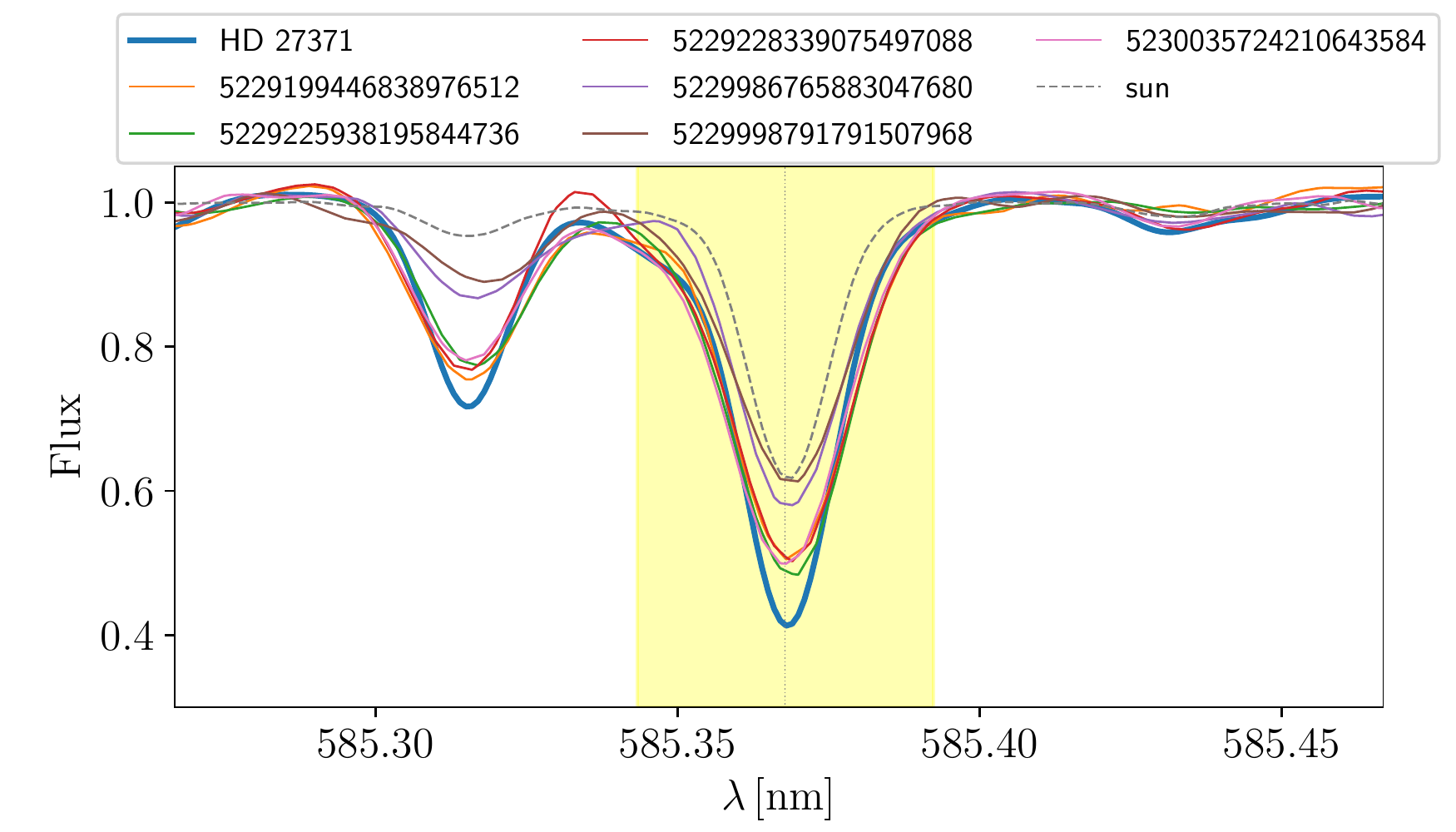}
\caption{Observed spectra for the \ion{Ba}{II} line at 585.3668 nm. In the thick blue line, the giant star from the Hyades, used as a reference for the differential analysis, and in the dashed grey line the Solar spectrum analysed from MIKE. The red clump giants from UBC~274 correspond to the orange, green, red and pink lines, and the subgiants are in brown and violet.}\label{fig:line}
\end{figure}

We notice that for certain elements the two sets of abundances can differ substantially and above the formal uncertainties, remarkably for Na, Mn, Y and Nd. This is in agreement with the results obtained in \citet{Casamiquela+2020}, where for the mentioned elements, the bracket abundances tend to have the largest trends with effective temperature, see their fig. 4. These trends are caused by systematic offsets in the absolute chemical abundance which depend on the spectral type. These offsets can be caused by a variety of factors such as non-LTE effects, systematics introduced by the analysis, or changes in chemical abundances in the stellar atmosphere along the stellar evolution.
The strictly line-by-line differential analysis helps to mitigate the mentioned systematic effects and therefore removes most trends with effective temperature, as shown by \citet[][fig 7]{Casamiquela+2020}. For it to be solid, this type of analysis needs a reference star as close as possible to the analysed star in terms of atmospheric parameters. The reference star from the Hyades is close to the four red clump stars of UBC~274 ($\sim100$ K in $\teff$ and $\sim0.3$ dex in $\logg$), but it is slightly different from the two subgiants ($\sim200$ K in $\teff$ and $\sim0.6$ dex in $\logg$). This difference is also clear visually comparing the spectra, for example, see the different depths of the \ion{Ba}{II} line in Fig.~\ref{fig:line} of the subgiants compared with the red clump stars and the reference star (HD~27371). These differences are most probably the cause of some small offsets in certain elements (e.g. Si, Ca, Ti, V, Fe) among the two groups of stars in Fig.~\ref{fig:chemab}. These offsets are always smaller than the mean quoted uncertainties, except for the case of Fe and Si, for which the two quantities are of the same order ($0.05$ dex).

We find that the uncertainties involved in the differential results are generally lower.
Because of this, and also for the previously mentioned reasons, we consider the differential abundances more accurate and precise than the absolute ones, and they will be used in the discussion.

\section{Discussion}\label{sec:discussion}

In the light of the outcomes of our global study of this cluster, including its morphology, kinematics and chemical abundances, we can discuss several of its properties.

\subsection{Cluster age and extinction}\label{sec:cluster}

\citet{Cantat-Gaudin+2020} reported a cluster age of $\log Age = 9.43$ (2.7 Gyr), distance modulus 11.25 mag ($d = 1778$ pc) and $A_V=0.24$. These values were derived massively for a large number of clusters using an artificial neural network, which takes into account \emph{Gaia} photometry and the mean parallax of the cluster. The algorithm was trained using a reference sample composed mainly of clusters from \citet{Bossini+2019}. 

Later, \citet{Piatti2020} found that the cluster is affected by differential reddening which ranges from $E_{B-V}=0.1\,\mhyphen\,0.25$. They found $\log Age = 9.45\pm0.05$ ($2.8\pm0.2$ Gyr) to provide the best fit to their CMD using the PARSEC isochrones \citep{Bressan+2012} with a compromise between the agreement of the turnoff and the red clump.
Both studies do not directly take into account the metallicity of the cluster to compute its age. In this study, we find that the cluster has a subsolar metallicity ($\sim-0.1$, see Sect.~\ref{sec:chemistry}), given the degeneration of the age and the metallicity, in this section, we re-derive its age. We also take into account the differential reddening for the isochrone fit.

\begin{figure}
\centering
\includegraphics[width=0.5\textwidth]{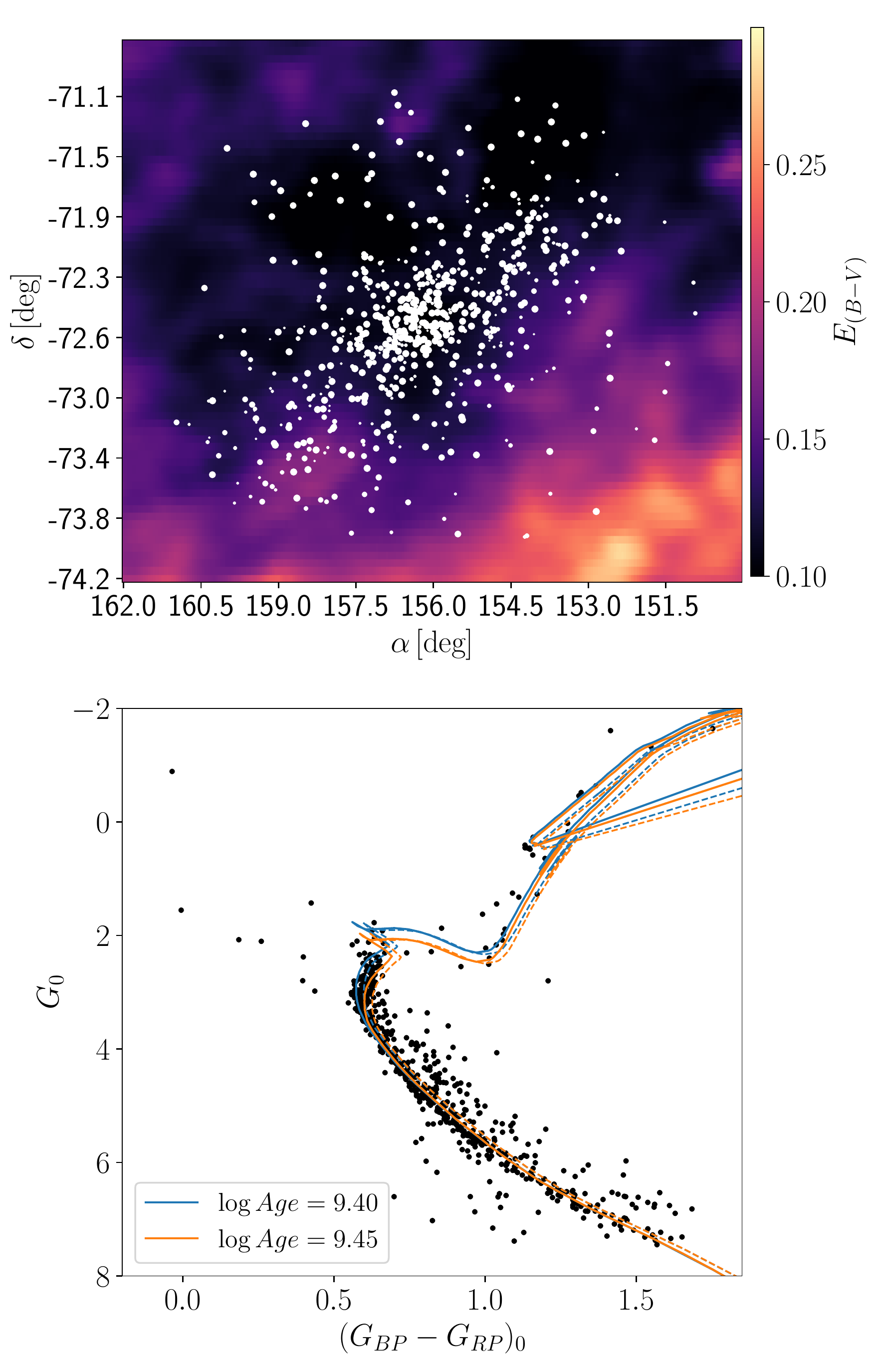}
\caption{Top: reddening 2D map of the cluster field with members overplotted (size is scaled with the probability of membership. Bottom: Intrinsic CMD in \gaiaedr3 passbands. Isochrones of two different ages (in blue and orange) and $\mh=-0.1$ are plotted with solid lines. The dashed lines correspond to the isochrones of $\mh=0.0$).}\label{fig:corrCMD}
\end{figure}

We plot in Fig.~\ref{fig:corrCMD} (top) the spatial distribution of $E_{(B-V)}$ in the cluster field. We use the \texttt{dustmaps} python package \citep{Green2018} to query the 2D dust extinction map of \citet{Schlafly+2011}, and we use their coefficients in table 6 to transform to physical reddening values $E_{(B-V)}$. We correct the magnitudes of each star for the corresponding reddening value, using the relative extinction values given by \citet{Wang+2019}. Additionally, we obtain the absolute magnitude $G_0$, shifting a distance modulus of 11.25 mag according to the cluster mean distance.
We plot in Fig.~\ref{fig:corrCMD} (bottom) the intrinsic CMD of the cluster members, and isochrones of two different ages accounting for a rather typical uncertainty of 10\% in age, and two different metallicities (solid and dashed lines representing [M/H]$=-0.1$ and 0.0, respectively). 

The intrinsic CMD effectively reduces the scatter along the main sequence and highlights a very remarkable equal-mass binary sequence, as analysed by \citet{Piatti2020}, but in a larger magnitude range. We notice some possible contaminants at the faint left side of the main sequence (corresponding to apparent magnitudes 17-19), in agreement with the studied recovery statistics where we can expect contamination of 30\% (Fig.~\ref{fig:classification}). At these magnitudes we are pushing the limits of both the data and the method, for a cluster located at almost 2 kpc, embedded in an extincted region and with proper motions highly mixed with the field distribution. Alternatively, these stars could also be members with less accurate photometry.
In the bright end, we find eight blue stragglers, which are recovered with a large membership probability. Six of them were already recovered in the list of \citet{Castro-Ginard+2020}, and two of them are newly identified in the present study.

It is visible from Fig.~\ref{fig:corrCMD} that the $\log Age=9.4$ isochrones (blue) reproduce better the turnoff, while the ones of metallicity $\mh=-0.1$ provide an adequate fit to the red clump position, which is shifted rightwards with the isochrones of Solar metallicity. Thus, taking into account the metallicity derived in this paper, we obtain an age of $\log Age=9.4$ (2.5 Gyr), slightly younger with respect to previous studies. This result highlights how the knowledge of the metallicity breaks the degeneracy and allows a more precise determination of the age, especially when we couple it with the high quality of the photometry and the precise mean parallax.

\subsection{Chemical properties}\label{sec:chemprop}
\begin{figure*}
\centering
\includegraphics[width=\textwidth]{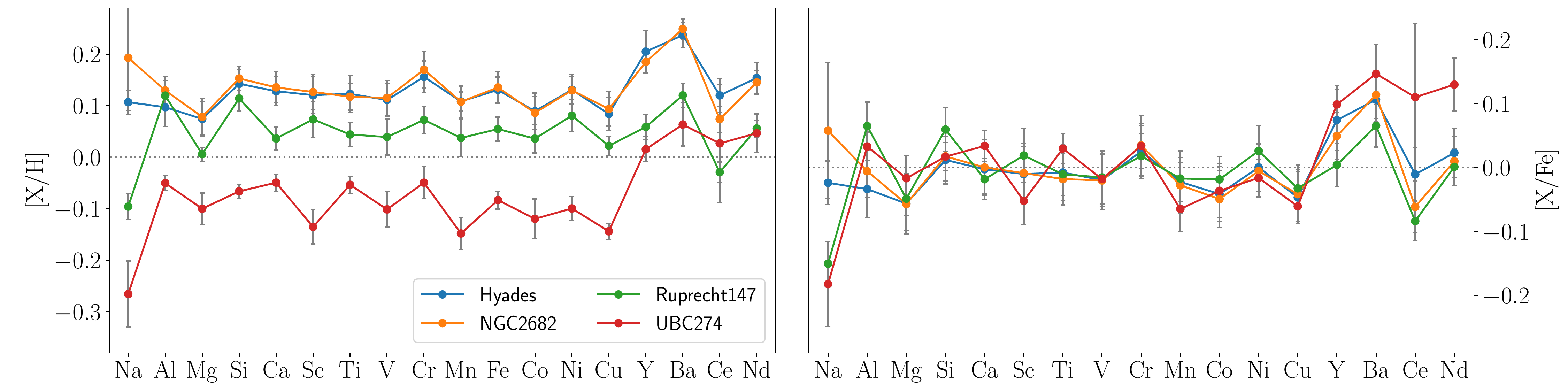}
\caption{Mean Solar-scaled differential abundances of UBC~274, in comparison with three other clusters studied in \citet{Casamiquela+2020}. Here we scaled the abundances of the three clusters to the Solar scale, so the plot is different wrt that in the original paper, where they were scaled to the Hyades abundance.}\label{fig:chemab_all}
\end{figure*}

We plot in Fig.~\ref{fig:chemab_all} the cluster average abundances of UBC~274 in comparison with the abundances of the giants of the Hyades, Praesepe and Ruprecht~147 analysed in \citet{Casamiquela+2020}. In this comparison, we have, on one hand, two intermediate-age ($\sim 650$ Myr) clusters in the immediate Solar neighbourhood (Hyades and Praesepe), which have an almost exact chemical signature. On the other hand, two old clusters with roughly the same age (2.5 and 3 Gyr), but in this case, Ruprecht~147 is in the Solar neighbourhood (280 pc from the Sun), while UBC~274 is located at 1780 pc from the Sun in the fourth Galactic quadrant. The mean abundance values computed for UBC~274 are listed in Table~\ref{tab:clusterabus}.

\begin{table}
\caption{Weighted mean abundances of UBC~274, obtained from the solar-scaled differential values.}\label{tab:clusterabus}
\centering
\setlength\tabcolsep{3pt}
\begin{tabular}{lrr}
 \hline
 Element & [X/H] & [X/Fe] \\
 \hline
Fe  & $-0.08 \pm 0.02$  & - \\
Na  & $-0.27 \pm 0.06$  & $-0.18 \pm 0.07$ \\
Al  & $-0.05 \pm 0.01$  & $ 0.03 \pm 0.02$ \\
Mg  & $-0.10 \pm 0.03$  & $-0.02 \pm 0.04$ \\
Si  & $-0.07 \pm 0.01$  & $ 0.02 \pm 0.02$ \\
Ca  & $-0.05 \pm 0.02$  & $ 0.03 \pm 0.02$ \\
Sc  & $-0.14 \pm 0.03$  & $-0.05 \pm 0.04$ \\
Ti  & $-0.05 \pm 0.02$  & $ 0.03 \pm 0.02$ \\
V & $-0.10 \pm 0.03$  & $-0.02 \pm 0.04$ \\
Cr  & $-0.05 \pm 0.03$  & $ 0.03 \pm 0.04$ \\
Mn  & $-0.15 \pm 0.03$  & $-0.06 \pm 0.04$ \\
Co  & $-0.12 \pm 0.04$  & $-0.04 \pm 0.04$ \\
Ni  & $-0.10 \pm 0.02$  & $-0.02 \pm 0.03$ \\
Cu  & $-0.14 \pm 0.02$  & $-0.06 \pm 0.02$ \\
Y & $ 0.02 \pm 0.02$  & $ 0.10 \pm 0.03$ \\
Ba  & $ 0.06 \pm 0.04$  & $ 0.15 \pm 0.05$ \\
Ce  & $ 0.03 \pm 0.11$  & $ 0.11 \pm 0.12$ \\
Nd  & $ 0.05 \pm 0.04$  & $ 0.13 \pm 0.04$ \\
\hline
\end{tabular}
\end{table}

From the left-hand panel in Fig.~\ref{fig:chemab_all}, and as seen also in the previous section, we find UBC~274 to be overall metal-poor ([Fe/H]$=-0.08\pm0.02$ dex). This contrasts with the slightly supersolar abundance of Ruprecht~147 ([Fe/H]$=0.05\pm0.02$), which has roughly the same age. This difference could be tentatively explained by the spatial distance between the two clusters, which is almost 2 kpc. However, the two clusters have similar Galactocentric radii: 7.8 kpc of UBC~274, to be compared with the 8.0 kpc of Ruprecht~147. 

Taking into account the overall decreasing radial metallicity trend, one would not expect a 2 Gyr cluster located basically at the Solar radius to be more metal-poor than the Sun. Using the gradient derived by \citep{Spina+2021}, with a slope of $-0.076\pm 0.009$ dex kpc$^{-1}$, a metallicity of -0.1 corresponds to a Galactocentric radius of 9.6 kpc. Taking the inferior margin of the nominal uncertainty in the linear fit, the metallicity and position of the cluster deviate 1.6$\sigma$ from the derived relation.
Several other authors have derived slightly shallower slopes, which would make the $\rgc$ prediction for this cluster even outer in the Galaxy, like \citet{Casamiquela+2019} $-$0.056$\pm$0.011~dex~kpc$^{-1}$ (18 open clusters), or \citep{Spina+2022} $-$0.064$\pm$0.007~dex~kpc$^{-1}$ (175 open clusters). But also others derived slightly steeper slopes such as \citet{Jacobson+2011} $-$0.10$\pm$0.02~dex~kpc$^{-1}$ (12 open clusters) or \citet{Netopil+2016} $-0.086\pm0.009$ (88 open clusters), which would make it more compatible with the cluster metallicity.

According to the orbit simulation done in Sect.~\ref{sec:morph} the cluster is nowadays close to its pericenter, and during the integration time, it sweeps a range of Galactocentric radii between 7.8 and 10.5 kpc. We did not attempt to integrate the orbit up to the cluster age because the reliability of the results decreases significantly with time due to inaccuracies in the time-dependence of the potential, amplification of the uncertainties in distance and motions, among other effects. Taking into account the kinematics and the chemical clues we speculate that the cluster is probably born outer in the Galaxy.

\subsubsection{Chemical pattern}
From the right-hand plot in Fig.~\ref{fig:chemab_all}, we see a marked depletion in Na abundance of UBC~274 with respect to the younger clusters, which is very consistent with the [Na/Fe] of Ruprecht~147, despite its different overall metallicity. This fact supports the explanation already stated in \citet{Casamiquela+2020}, where they saw that this Na depletion is only detected for the giants in Ruprecht~147, and not its F and G dwarfs. They attributed this to an effect of internal mixing in the surface of massive red clump giants (younger clusters, like the Hyades), which can present overabundances of Na of 0.2-0.3 dex, with respect to its dwarfs. Thus, our abundances of Na are not representative of the true cluster abundances, since the transformation from the differential to Solar-scaled abundances assumes a homogeneous abundance of the reference cluster (the Hyades), and for Na, this would not be the case. This effect is also analysed in detail in other works such as \citet{Smiljanic+2018}.

UBC~274 has a compatible signature with that of Ruprecht~147 for most of the analysed elements with the remarkable exception of the s-process elements. The two clusters differ around 0.1 dex in Y, Ba and Nd, which is above the uncertainties (around 0.04 dex). Ce displays larger uncertainties in UBC~274 (0.1 dex) so the difference is not significant.
We speculate that this could be due to the difference in metallicity between the two clusters. According to several works \citep[e.g.][]{Karakas+2016} there is a strong metallicity dependence of stellar yields, particularly of s-process elements. These elements are secondary elements formed through neutron captures and thus depend on the number of iron seeds available and the neutron flux during the AGB phase. Additionally, at super solar metallicity, the thermal pulses of AGB stars are reduced. Overall, the models indicate that at higher metallicity there is a lower expected abundance of s-process elements. The models of \citet{Karakas+2016} obtain significantly stronger production from subsolar ($\feh\sim-0.3$ dex) vs Solar AGB stars. 
In our case, the difference in metallicity among the two clusters is smaller (0.15 dex), but given the high precision in abundances, we are also able to detect a small difference in s-process abundance (0.1 dex).

\begin{figure}
\centering
\includegraphics[width=0.5\textwidth]{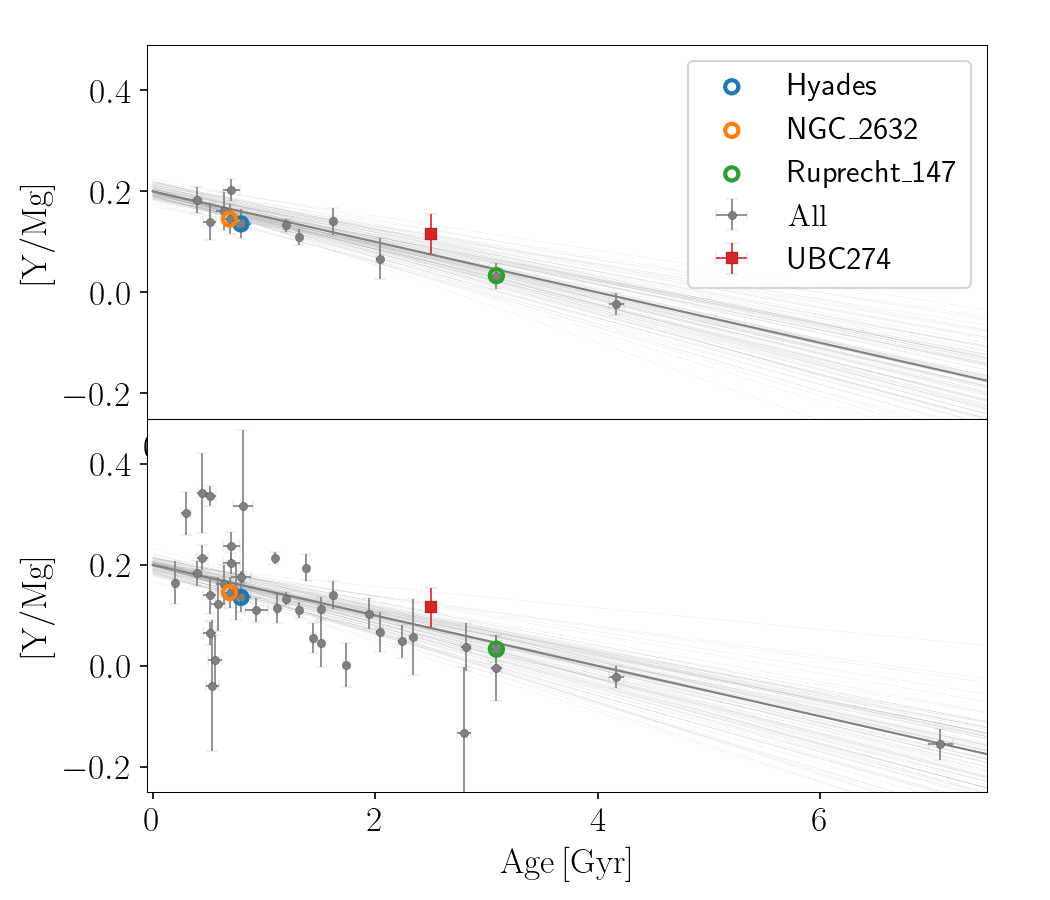}
\caption{[Y/Mg] abundances of the sample of clusters from \citet{Casamiquela+2021} in grey, and the linear fit computed. The values of the Hyades, NGC~2632 and Ruprecht~147 are highlighted. The value computed in this paper for UBC~274 is plotted as a red square. The top plot shows the clusters closer than $d<1$ kpc, and the bottom plot all the clusters in the sample.}\label{fig:YMg}
\end{figure}

Observations of open clusters from the Gaia-ESO survey \citep{Casali+2020} also show an indication of metallicity dependence of the s-process yields, where the clusters at $\feh\sim0.2$ dex exhibited large discrepancies in [Y/Mg] compared to those at Solar and subsolar metallicities. We computed the [Y/Mg] ratio for this cluster and we compare it with the sample of clusters in \citet{Casamiquela+2021} in Fig.~\ref{fig:YMg}. Its value and uncertainty make UBC~274 compatible with the overall trend with age computed by \citet{Casamiquela+2021}, which is similar to that computed by other previous studies \citep[e.g.][]{Spina+2018,Jofre+2019}. This means that even though the [Y/Fe] abundance of this cluster is enhanced, the [Mg/Fe] value is proportionally higher than in the case of Ruprecht~147, and compensates for the [Y/Mg] ratio. 

The compatibility of the [Y/Mg] clock for this cluster indicates that it has had a similar enrichment history as the other clusters in the solar neighbourhood, even though it is quite distant and has a different metallicity.
However, understanding the full picture is complicated because possibly the interstellar medium enrichment also has spatial dependencies, not only as a function of metallicity, as discussed in \citet{Casamiquela+2021}.

\subsubsection{Chemical homogeneity}
The weighted mean values of the cluster abundances obtained from the differential analysis (Table~\ref{tab:clusterabus}) show a high consistency among the abundances of all analysed stars. Standard deviations for most elements (exception of Na and Ce) are of the order of 0.03 dex.

Since our sample is small in terms of size we cannot put strong constraints on the overall chemical homogeneity of the cluster. However, the spectroscopic targets analysed are not fully concentrated in the central part of the projected sky distribution of the cluster (see Fig.~\ref{fig:targets}). They cover an area of around $\sim1.5$ deg, which represents a physical separation of $\sim35$ pc at the cluster distance. In this region of the cluster, we see no evidence of chemical inhomogeneity, at least traced by giant stars.

Nonetheless, our spectroscopic targets do not expand the full extension of the tidal structure that is identified (100 pc). This is possibly due to mass segregation (Sect.~\ref{sec:segregation}), which made it difficult to find giants further away from the centre. Observations of dwarfs covering the tidal structure would be of interest to better assess the chemical homogeneity of the cluster's members.

Chemical homogeneity has been also evaluated in several recently discovered "strings", kinematically identified stellar groups \citep[][]{Kounkel+2019} which extend from 80 to 300 pc. \citet{Manea+2022} studied 10 of these elongated structures using GALAH abundances finding that most of them show a higher chemical homogeneity than the local field, and similar abundance scatters to most studied OCs. This would support a co-natal origin for these chemically and kinematically related groups of stars. They present two scenarios to explain their origin: i) strings are born at smaller spatial scales ($\sim$ 5 pc) and elongated afterwards; or ii) star formation occurs at larger spatial scales, then implying that the interstellar medium is more chemically homogeneous at those spatial scales than previously suggested. The tidal structure of UBC~274 expands similar spatial scales as the strings and agrees with our simulation in Sect.~\ref{sec:morph} which starts with a highly concentrated cluster in the past (5 pc). Thus, in this case, a chemically homogeneous tidal structure seems to favour the first hypothesis. UBC~274 is, however, older than the average age of the stellar groups analysed in \citet{Manea+2022}, therefore for this particular case, there has been more time to tidally elongate the cluster. Also, our sampled area is restricted for methodological reasons, even though our simulation indicates that the tidal extent of the cluster is much larger. Chemical studies of recently discovered tidal tails in several young/old open clusters \citep[][]{Tarricq+2021b}, coupled with simulations of the tidal disruption of open clusters could shed more light on this matter.

\section{Summary and conclusions}\label{sec:conclusions}
The recent discovery of a significant number of open clusters with spatially elongated morphology presents an opportunity to study the disruption mechanisms of co-natal stars, and how they populate the Galactic field. The increasing quality of \emph{Gaia} data releases allows a better-constrained membership determination of these clusters and up to fainter magnitudes. Precise chemical abundances of stars covering the tidal region provide an insight into the chemical homogeneity of co-natal stars.
UBC~274 is a recently discovered OC \citep{Castro-Ginard+2020}, and is one of the oldest (2.5 Gyr) and further away (1778 pc) clusters with the presence of a remarkable tidal tail. We perform a morphological, kinematic and chemical study of this cluster to study its disruption.

We use \gaiaedr3 data to extend the membership list of this cluster up to magnitude $G=19$ and cover a distance of 50 pc around the cluster centre. We fit a Gaussian Mixture Model on the projected sky distribution, where one of the components presents an eccentricity of 0.93 and tilted 10 deg with respect to the Galactic plane. The elongated structure and its tilt are fully compatible with the results of a test particle simulation of a cluster mass loss. The simulation predicts a much larger tidal structure outside of our area of analysis, and it indicates that we are observing the internal S shape of the tidal stream.

We analyse mass segregation in this cluster by comparing the length of the MST of massive stars with that of random stars, as done in \citet{Tarricq+2021b}. We find a significant degree of mass segregation: the $N=$10 most massive stars ($G<12$) appear to be 1.5 times closer than 10 randomly chosen stars. This difference quickly vanishes when we compute the MST length ratio for $N>15$, where no net mass segregation is obtained.
This contrasts with previous findings of \citet{Piatti2020}, who found no evidence of mass segregation in this cluster comparing the distribution of binaries with non-binaries. The different types of stars used in the two studies is most likely the cause of this difference since we find mass segregation of intrinsically more massive stars with respect to low-mass stars.

We obtained high-resolution and high S/N spectra of eight stars of this cluster using the MIKE spectrograph at the 6.5m Clay Magellan Telescope at Las Campanas Observatory. We are able to compute chemical abundances of 18 elements for four red clump stars and two subgiants, following a differential line-by-line approach with respect to a red clump star in the Hyades. The cluster shows an overall subsolar metallicity [Fe/H]$=-0.08\pm0.02$, which is not fully consistent with the Galactocentric radius where it is located (7.8 kpc) taking into account a rather typical value of a decreasing metallicity gradient. Since our orbit integration shows that the cluster sweeps Galactocentric distances between 7.8 and 10.5 kpc, we speculate that the cluster could have been born slightly outer in the Galaxy. UBC~274 shows a similar chemical pattern to the cluster of similar age Ruprecht~147 (located at 280 pc from the Sun) analysed in \citet{Casamiquela+2020}, with the exception of neutron-capture elements.
We find UBC~274 to be overabundant in [Y/Fe], [Ba/Fe] and [Nd/Fe] by 0.1 dex, above the quoted uncertainties. This difference can be explained by the dependence of s-process stellar yields with the metallicity, given the difference in the metallicity of the two clusters. However, the agreement of the [Y/Mg] values of the cluster with the expected relation with age, indicates that the cluster has had a similar enrichment history as the solar neighbourhood.

From our spectroscopic targets, covering a physical area of 35 pc, we find a remarkably high chemical homogeneity with abundance dispersions of the order of 0.03 dex. Since the number of targets is small we cannot put strong constraints on the overall chemical homogeneity. A larger number of spectroscopic observations covering the full region of the tidal structures of a large number of elongated clusters recently discovered \citep{Tarricq+2021b} would give a better insight into the timescales and interplay between the star formation process and cluster disruption.

\begin{acknowledgements}
This work has made use of data from the European Space Agency (ESA) mission \emph{Gaia} (\url{http://www.cosmos.esa.int/gaia}), processed by the \emph{Gaia} Data Processing and Analysis Consortium (DPAC, \url{http://www.cosmos.esa.int/web/gaia/dpac/consortium}). We acknowledge the \emph{Gaia} Project Scientist Support Team and the \emph{Gaia} DPAC. Funding for the DPAC has been provided by national institutions, in particular, the institutions participating in the \emph{Gaia} Multilateral Agreement.
This research made extensive use of the SIMBAD database, and the VizieR catalog access tool operated at the CDS, Strasbourg, France, and of NASA Astrophysics Data System Bibliographic Services.

We acknowledge support from "programme national de physique stellaire" (PNPS) and from the "programme national cosmologie et galaxies" (PNCG) of CNRS/INSU. J.O. acknowledges financial support from the Agencia Estatal de Investigación of the Ministerio de Ciencia, Innovación y Universidades through project PID2019-109522GB-C53. We acknowledge the support form the ECOS Sud-CONICYT exchange program number 180049.

\end{acknowledgements}

\bibliographystyle{aa} 
\bibliography{biblio2_v2}

\begin{appendix}
\section{Membership statistics}\label{sec:appendix}
We chose the most optimal threshold in the probability of membership, testing the outcomes of our clustering method in a synthetic cluster. The cluster and field properties and uncertainties were generated as explained in Sect.~\ref{sec:membership}

\begin{figure}[h!]
\centering
\includegraphics[width=0.45\textwidth]{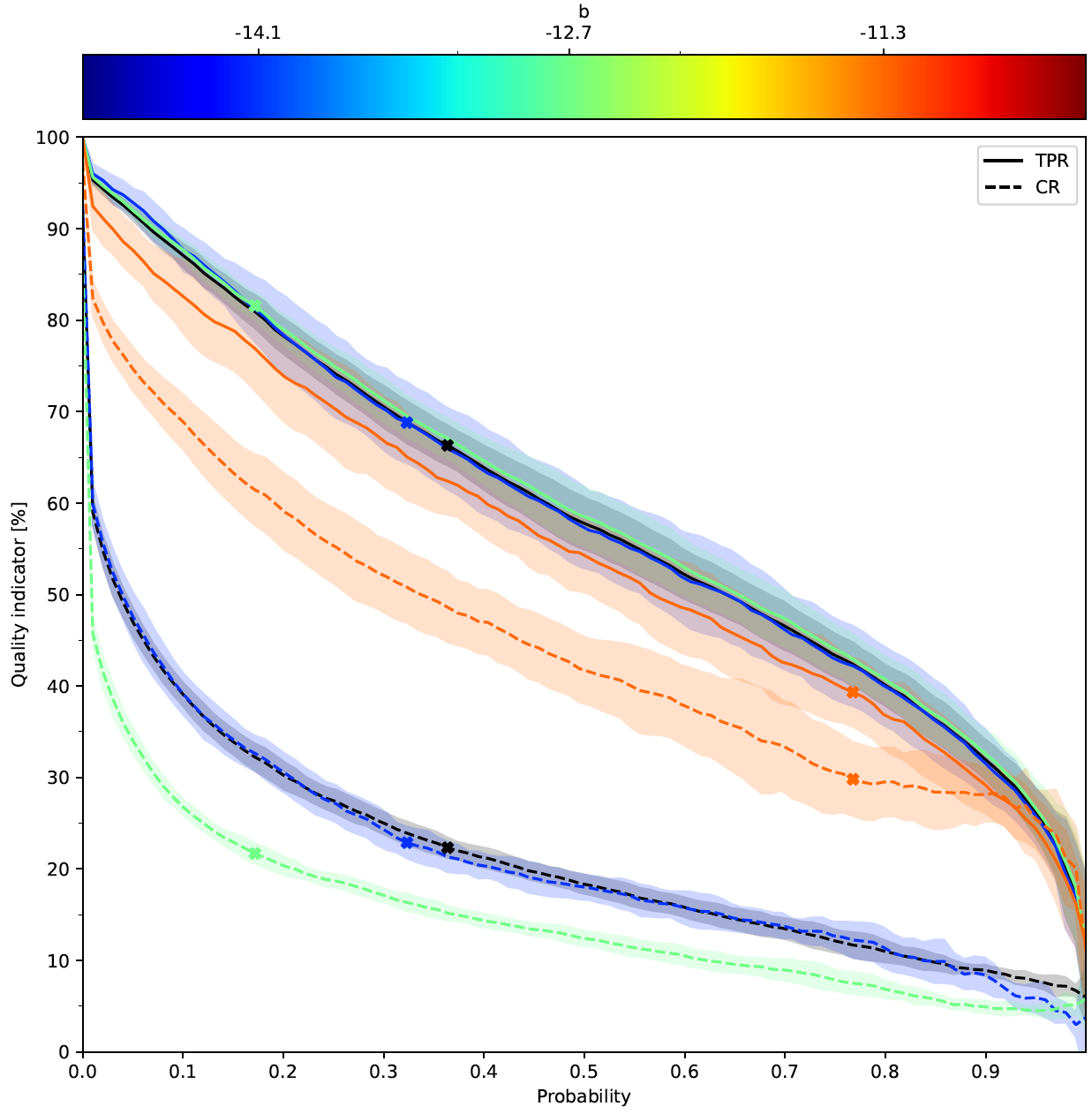}
\includegraphics[width=0.45\textwidth]{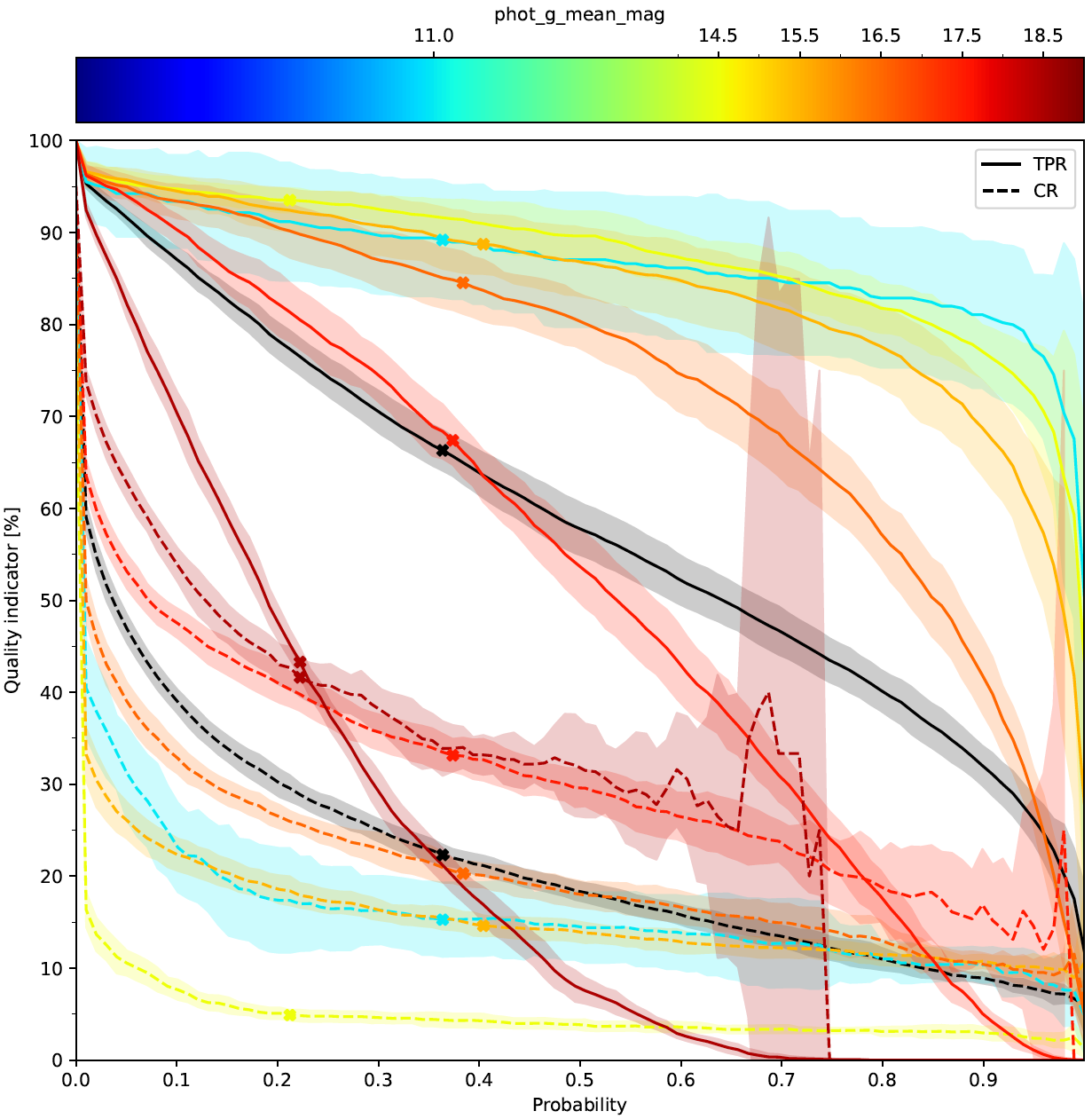}
\caption{TPR (solid lines) and CR (dashed lines) as a function of the probability computed by our membership method (Sect.~\ref{sec:memb}), in bins of Galactic latitude $b$ (top) and $G$ magnitude (bottom). The crosses indicate the optimum probability thresholds that maximize the accuracy of the classifier (i.e., the fraction of correct classifications) at each bin, and are the ones detailed in Table~\ref{tab:thresholds}. The black line indicates the global performance, and the shaded regions depict the standard deviation of the ten synthetic simulations.}\label{fig:classification}
\end{figure}

We test the true positive rate (TPR) and the contamination rate (CR) sweeping different cuts in the probability threshold in Fig~\ref{fig:classification}. In the top panel, we show the TRP and CR curves of three bins of Galactic latitude $b$, and the bottom panel in six bins of $G$ magnitude. For each bin, we find the probability value that provides the lowest CR and the highest TPR (marked with a cross in the plots). We recall the reader that:
\begin{equation}
    TPR = \frac{TP}{TP + FN}, \\
    CR = \frac{FP}{FP + TP}
\end{equation}
where TP is the number of true positives, FN is the number of false negatives, and FP is the number of false positives.

According to our simulation, we can expect recovery rates better than 85\% for sources with $G<17$, and this goes down to 67\% and 43\% for sources $17<G<18$ and $18<G<19$, respectively. We see a clear difference as a function of latitude: we expect better recovery statistics in the central bin with true positive rates of around 80\% and contamination of around 20\%. In the two external bins of the cluster the probability thresholds are larger to minimise the contamination rates, particularly in the northern part (closer to the Galactic plane), where we obtain a true positive rate of 40\% with expected contamination of 30\%.

\begin{table}[h!]
\caption{Optimal probability thresholds were obtained by bins of $G$ magnitude (top) and of Galactic latitude $b$ (bottom). We include the total number of sources of the simulation in each bin, the true positive rate (TPR) and the contamination rate (CR).}\label{tab:thresholds}
\centering
\begin{tabular}{lcccc}
 \hline
 \multicolumn{5}{c}{$G$ [mag]} \\
 \hline
 \hline
 bin   & Probability & Num     & TPR  & CR \\
       & threshold   & sources & (\%) & (\%) \\
 \hline
 8-14  & 0.36        & 461  & 89 & 15\\
 14-15 & 0.21        & 640  & 94 & 5 \\
 15-16 & 0.40        & 1309 & 89 & 15\\
 16-17 & 0.38        & 2726 & 85 & 20\\
 17-18 & 0.31        & 4666 & 67 & 33\\
 18-19 & 0.22        & 6951 & 43 & 42\\
 \hline
\end{tabular}

\vspace{0.5cm}

\begin{tabular}{lcccc}
 \hline
 \multicolumn{5}{c}{$b$ [deg]} \\
 \hline
 \hline
 bin   & Probability & Num     & TPR  & CR \\
       & threshold   & sources & (\%) & (\%) \\
 \hline
 $-15<b<-13.2$   & 0.32        & 3698  & 69 & 23 \\
 $-13.2<b<-12.2$ & 0.17        & 6687  & 82 & 22 \\
 $-12.2<b<-11$.  & 0.77        & 6372  & 39 & 30 \\
 \hline
\end{tabular}
\end{table}

\end{appendix}

\end{document}